\documentclass[prd,aps,nofootinbib,twocolumn,superscriptaddress,preprintnumbers,balancelastpage,longbibliography]{revtex4-1}

\usepackage{ae,aecompl}
\usepackage{graphicx}	
\usepackage{amsmath}	
\usepackage{graphicx}
\usepackage{dcolumn}
\usepackage{bm}
\usepackage{graphics}
\usepackage{afterpage}
\usepackage{float}
\usepackage{subfigure}
\usepackage{rotating}
\usepackage{multirow}
\usepackage{tabularx}
\usepackage{booktabs}
\usepackage{multirow}
\usepackage{fancyheadings}
\usepackage{theorem}
\usepackage{moreverb}
\usepackage{euscript}
\usepackage{psfrag}
\usepackage{slashed}
\usepackage{mathtools}
\usepackage{makecell}

\usepackage[colorlinks=true
,urlcolor=blue
,anchorcolor=blue
,citecolor=blue
,filecolor=blue
,linkcolor=red
,menucolor=blue
,linktocpage=true
,pdfproducer=medialab
,pdfa=true
]{hyperref}
\usepackage{listings}
\usepackage{color,xcolor}

\newcolumntype{C}[1]{>{\centering\let\newline\\\arraybackslash\hspace{0pt}}m{#1}}

\newcommand\Tstrut{\rule{0pt}{2.6ex}}         
\newcommand\Bstrut{\rule[-0.9ex]{0pt}{0pt}}   

\definecolor{deepblue}{rgb}{0,0,0.9}
\definecolor{deepred}{rgb}{0.85,0,0}
\definecolor{deepgreen}{rgb}{0,0.95,0}

\lstdefinestyle{python}{
  belowcaptionskip=1\baselineskip,
  breaklines=true,
  frame=L,
  xleftmargin=\parindent,
  language=Python,
  showstringspaces=false,
  basicstyle=\small\ttfamily,
  morekeywords={models, lambda, forms,True,False,None},
  keywordstyle=\bfseries\color{deepgreen!40!black},
  commentstyle=\itshape\color{gray},
  identifierstyle=\color{black},
  stringstyle=\color{deepred},
  rulecolor=\color{gray},
}

\newcommand{\es}[2] {\begin{equation} \label{#1} \begin{split} #2 \end{split} \end{equation}}
\newcommand{\be}{\begin{equation}}
\newcommand{\ee}{\end{equation}}

\begin{document}

\title{Foreground Mismodeling and the Point Source Explanation \\
of the {\it Fermi} Galactic Center Excess}
\author{Malte Buschmann}
\affiliation{Leinweber Center for Theoretical Physics, Department of Physics, University of Michigan, Ann Arbor, MI 48109 USA}
\author{Nicholas L. Rodd}
\affiliation{Berkeley Center for Theoretical Physics, University of California, Berkeley, CA 94720, USA}
\affiliation{Theoretical Physics Group, Lawrence Berkeley National Laboratory, Berkeley, CA 94720, USA}
\author{Benjamin R. Safdi}
\affiliation{Leinweber Center for Theoretical Physics, Department of Physics, University of Michigan, Ann Arbor, MI 48109 USA}
\author{Laura J. Chang}
\affiliation{Department of Physics, Princeton University, Princeton, NJ 08544, USA}
\author{Siddharth Mishra-Sharma}
\affiliation{Center for Cosmology and Particle Physics, Department of Physics, New York University, New York, NY 10003, USA}
\author{Mariangela Lisanti}
\affiliation{Department of Physics, Princeton University, Princeton, NJ 08544, USA}
\author{Oscar Macias}
\affiliation{Kavli Institute for the Physics and Mathematics of the Universe (WPI), University of Tokyo, Kashiwa, Chiba 277-8583, Japan}
\affiliation{GRAPPA Institute, University of Amsterdam, 1098 XH Amsterdam, The Netherlands}

\preprint{LCTP-20-02}
\date{\today}

\begin{abstract}
The {\it Fermi} Large Area Telescope has observed an excess of $\sim$GeV energy gamma rays from the center of the Milky Way, which may arise from near-thermal dark matter annihilation.  Firmly establishing the dark matter origin for this excess is however complicated by challenges in modeling diffuse cosmic-ray foregrounds as well as unresolved astrophysical sources, such as millisecond pulsars.  Non-Poissonian Template Fitting~(NPTF) is one statistical technique that has previously been used to show that at least some fraction of the GeV~excess is likely due to a population of dim point sources.  These results were recently called into question by Leane~and~Slatyer~(2019), who showed that a synthetic dark matter annihilation signal injected on top of the real {\it Fermi} data is not recovered by the NPTF procedure.  In this work, we perform a dedicated study of the {\it Fermi} data and explicitly show that the central result of Leane~and~Slatyer~(2019) is likely driven by the fact that their choice of model for the Galactic foreground emission does not provide a sufficiently good description of the data. 
We repeat the NPTF analyses using a state-of-the-art model for diffuse gamma-ray emission in the Milky Way and introduce a novel statistical procedure, based on spherical-harmonic marginalization, to provide an improved description of the Galactic diffuse emission in a data-driven fashion.  With these improvements, we find that the NPTF results continue to robustly favor the interpretation that the Galactic Center excess is due, in part, to unresolved astrophysical point sources across the analysis variations that we have explored.
\end{abstract}

\maketitle

\section{Introduction}

The {\it Fermi} Galactic Center Excess (GCE) is an approximately spherically symmetric excess of $\sim$GeV gamma-rays observed in the inner regions of the Milky Way by the {\it Fermi} Large Area Telescope (LAT).  While the GCE is subdominant compared to diffuse cosmic-ray emission in this region of sky, the statistical and systematic robustness of the excess to variations in dataset and foreground models has been firmly established~\cite{Goodenough:2009gk,Hooper:2010mq,Boyarsky:2010dr,Hooper:2011ti,Abazajian:2012pn,Hooper:2013rwa,Gordon:2013vta,Abazajian:2014fta,Daylan:2014rsa,Calore:2014xka,Abazajian:2014hsa,TheFermi-LAT:2015kwa,Linden:2016rcf,Macias:2016nev,Clark:2016mbb}.  The GCE has attracted significant attention because it may arise from the annihilation of a near-thermal dark matter~(DM) candidate with mass on the order of $\sim$10--100 GeV.  Furthermore, the spatial morphology of the GCE is consistent with that expected from annihilating DM following a generalized Navarro-Frenk-White~(NFW) density profile~\cite{Navarro:1995iw,Navarro:1996gj}.  However, there are claims that the photon-count statistics of the GCE are more consistent with the excess arising in part from a population of sub-threshold (\emph{i.e.}, not individually resolvable) astrophysical point sources (PSs) and not DM annihilation, the latter of which would be smoothly distributed in the Inner Galaxy~\cite{Lee:2015fea,Bartels:2015aea}.  Sub-threshold PSs are expected in the Inner Galaxy, and millisecond pulsars in particular could possess an energy spectrum consistent with that observed for the GCE and may also be distributed spatially in such a way as to explain the observed morphology of the GCE~\cite{Abazajian:2014fta,Abazajian:2010zy,Hooper:2013nhl,Calore:2014oga,Cholis:2014lta,Petrovic:2014xra,Yuan:2014yda,OLeary:2015gfa,Brandt:2015ula}. Indeed, recent studies have suggested the GCE is correlated with stellar overdensities in the Inner Galaxy~\cite{Macias:2016nev,Macias:2019omb,Bartels:2017vsx}.

In this paper, we examine the extent to which mismodeling Galactic foreground emission may bias the evidence for a PS explanation of the GCE and propose methods for mitigating such effects.  We focus specifically on the Non-Poissonian Template Fitting~(NPTF) analysis framework used in Ref.~\cite{Lee:2015fea} to provide evidence for unresolved PSs in the Inner Galaxy.  The NPTF was developed in Refs.~\cite{Lee:2014mza, Lee:2015fea}, expanding upon earlier applications of one-point fluctuation analyses to gamma rays~\cite{Miyaji:2001dp,Malyshev:2011zi}.  The central idea behind the NPTF is that sub-threshold PSs that are not modeled explicitly manifest as non-Poissonian fluctuations over the background expectation in pixelated data.  The NPTF is based upon a likelihood function framework that includes  Poissonian templates that describe smooth emission processes, whose spatial morphology is known, and non-Poissonian templates that describe the probabilistic distribution of PSs on the sky (whose exact positions are unknown) and their luminosity function. We examine how mismodeling the foreground emission described by the smooth Poissonian templates may affect inferences about the non-Poissonian (PS-like) components when applied to the GCE.  

The dominant source of gamma-ray flux in the Inner Galaxy of the Milky Way arises from diffuse emission due to the interactions of cosmic rays with interstellar gas and radiation.  For example,  high-energy protons can scatter inelastically with gas, producing pions that decay into photons.  Bremsstrahlung emission from cosmic-ray electrons scattering off of the same gas is also important.  Both of these sources of emission trace the gas distribution in the Milky Way, modulated by the density of cosmic rays, and thus exhibit structure on small angular scales.  An additional source of diffuse emission arises from the inverse Compton~(IC) process of cosmic-ray electrons up-scattering the interstellar radiation field.  This process does not trace the gas distribution and does not have structure on small angular scales. Diffuse mismodeling can affect many aspects of the reconstruction of the GCE, including its energy distribution and spatial morphology---see \emph{e.g.}, Refs.~\cite{Calore:2014nla,Carlson:2015ona,Macias:2016nev}.  It also can have an impact when studying the PS nature of the GCE.  For example, one  worry is that mismodeling the gas-correlated diffuse emission can generate artificial structures on small angular scales, since the diffuse emission has a small-scale component arising from the gas distribution, and this mismodeled emission can be incorrectly interpreted as arising from astrophysical PSs~\cite{Lee:2015fea,Leane:2019uhc}.  Mismodeling of the IC component can also be  problematic, because large-scale residuals may be spuriously interpreted as a population of PSs, especially given that the NPTF does not use any information on the spatial correlations of residuals.  

The evidence in favor of a PS explanation of the GCE has been recently questioned by Leane~and~Slatyer~\cite{Leane:2019uhc}, which claimed that the NPTF results for the GCE are not self-consistent in that a synthetic DM annihilation signal injected on top of the true {\it Fermi} data is not correctly recovered.  In our companion paper~\cite{Chang:2019ars}, we performed such signal injection tests on simulated data and cautioned that the  results need to be interpreted with great care.  We showed that even in the pure Monte Carlo (MC) setting where the underlying emission components are perfectly modeled, the correct injected signal flux may not be recovered properly due to biases induced by the NPTF priors and the inherent degeneracy between emission from a population of ultrafaint PSs and truly smooth emission. Additionally, we demonstrated that these challenges are further exacerbated by issues with diffuse mismodeling, which are certainly present in the real data. 

In this work, we focus on analyzing the interplay between diffuse mismodeling and evidence for PSs in the real {\it Fermi} data. To minimize the biases arising from the fundamental degeneracy between ultrafaint PSs and DM, we restrict our study to sources that are bright enough to be distinguishable from DM, but which still fall below {\it Fermi}'s threshold to be resolved as individual PSs.  We perform a careful treatment of the diffuse emission modeling, following three different approaches.  First, we construct improved diffuse emission templates, closely related to those used in Refs.~\cite{Macias:2016nev,Macias:2019omb}, which provide a substantially improved fit to the data.  Second, we propose a novel technique for mitigating mismodeling whereby we perform a spherical-harmonic decomposition of the diffuse foreground model skymap, treating the low-$\ell$ spherical-harmonic coefficients (describing large-scale structures) as nuisance parameters.
By marginalizing over the large-scale variations, we can correct for possible mismodeling effects in a data-driven way, without adding additional degrees of freedom on small angular scales.  Lastly, we consider the effect of shrinking the size of the region of interest (ROI) in order to mitigate large-scale mismodeling issues.  

From the tests that we perform on the {\it Fermi} data in this paper, we can conclude the following:
\begin{itemize}
    \item The results of the signal injection tests performed by Leane~and~Slatyer~\cite{Leane:2019uhc} are due to mismodeling the Milky Way diffuse emission.  Repeating these tests with improved foreground models, we find that artificial DM signals injected on the {\it Fermi} data are correctly recovered by the NPTF.
    \item While diffuse mismodeling likely affected the original NPTF analysis in Ref.~\cite{Lee:2015fea}, the evidence for spherical PSs in the Inner Galaxy is \emph{robust} to the variations we have tested, even after mitigating the effects of diffuse mismodeling. 
\end{itemize}

The remainder of this work is organized as follows.  We start by providing a brief summary of the methods and models used in the analyses.  Next, we show that standard diffuse models used in the literature suffer from over-subtraction even at the level of Poissonian template fits in the Inner Galaxy, while our improved diffuse model, along with other more up-to-date diffuse models, does not.  We also show that the spherical-harmonic marginalization procedure is effective, at the level of Poissonian template fits, at mitigating over-subtraction.    
Next, we present results for the NPTF in the Inner Galaxy using \emph{(i)}~updated diffuse models, and \emph{(ii)}~spherical-harmonic marginalization, and then we consider how the results depend on the size of the ROI.  Additional results are presented in the Appendices, such as a discussion of the new 4FGL PS mask~\cite{Fermi-LAT:2019yla} in App.~\ref{app:4FGL}, results without any PS mask in App.~\ref{app:unmasked}, an analysis of the absolute goodness of fit of the diffuse models in App.~\ref{app:qual}, and results with a novel high-resolution gas template in App.~\ref{app:HI}.

\section{Analysis Methods}

In this work, our goal is to probe the PS nature of the GCE in a manner that reduces the potential confusion between smooth emission and dim PSs and minimizes the impact of imperfect Galactic diffuse models.
In this section, we outline the tools and dataset that we use to achieve this aim.
To begin, we provide a brief review of the NPTF method itself and the dataset used.
Then, we turn to an overview of the suite of Galactic diffuse emission models considered in this work, with a particular emphasis on the more recent hydrodynamical model employed.
This section concludes with a description of the novel harmonic marginalization procedure that we introduce as a way to marginalize over large-scale uncertainties in the diffuse emission, without impacting the small-scale structure that is the hallmark of unresolved PSs.

\subsection{Non-Poissonian Template Fitting}

To test for the presence of PSs within the {\it Fermi} gamma-ray data, we use the NPTF method, which was first developed in Refs.~\cite{Malyshev:2011zi,Lee:2014mza,Lee:2015fea}.\footnote{Specifically, we use the publicly available \href{https://github.com/bsafdi/NPTFit/}{\texttt{NPTFit}} code~\cite{Mishra-Sharma:2016gis}.}
The NPTF is a generalization of the conventional astrophysical template fitting approach, which describes a photon dataset as a Poisson draw from a linear combination of sky maps, where each map is associated with a particular source of gamma-ray emission.
In more detail, if we pixelate the dataset in a single energy bin so that it is represented as a list of integers $\{ n_p \}$, with $n_p$ the number of counts in each pixel $p$, then the data is modeled as a set of templates $T_p^t$.\footnote{All sky maps are pixelated according to \texttt{HEALPix}~\cite{Gorski:2004by,Zonca2019}, taking \texttt{nside}=128.}
Here, $t$ indexes the different templates, which are given arbitrary normalization.
In the conventional approach, the expected number of counts in each pixel is $\mu_p({\bm \theta})= \sum_t A_t T_p^t$, where the model parameters ${\bm \theta} = \{ A_t \}$ are just the individual template normalizations.
These normalizations are inferred from the data through the use of a Poisson likelihood.
For a given model $\mathcal{M}$ that specifies a set of templates, the likelihood function is
\es{eq:likelihood}{
p(d |{\bm \theta}, \mathcal{M}) = \prod_p p_{n_p}^{(p)}({\bm \theta}) \,,
}
where $d = \{ n_p \}$, and the individual probabilities are given by the Poisson distribution
\be
p_{n_p}^{(p)}({\bm \theta}) = \frac{\mu_p^{n_p}({\bm \theta})}{n_p!} e^{-\mu_p({\bm \theta})}\,.
\ee

The NPTF generalizes the above formalism to incorporate templates that trace the spatial distribution of unresolved PSs and account for their presence statistically.  We note that resolved PSs whose spatial locations are known may be modeled directly using Poissonian templates. 
The challenge is then to go from a map of the distribution of unresolved PSs to the probability of observing a given number of counts in a pixel.  The procedure for doing so is controlled by the following three processes.
First, we need the probability of a given number of PSs in a pixel, which is controlled by the Poisson distribution with mean set by the expected number of sources.
Next, for each source, we draw the expected number of counts from a source-count distribution, described in detail below.
Finally, we determine the actual number of counts for each source, which is controlled by the Poisson distribution with mean set to the expected number of counts for each source.
Beyond these three steps, there are several technical details that must be accounted for, such as the effect of the finite point-spread function~(PSF) of the instrument.
Ultimately, however, all these factors can be incorporated and a modified $p_{n_p}^{(p)}({\bm \theta})$ derived, allowing for a non-Poissonian version of the likelihood in Eq.~\eqref{eq:likelihood}.
We eschew the details from the present discussion, and refer to~\cite{Mishra-Sharma:2016gis} for an extensive review.

The central ingredient of the non-Poissonian model is the source-count distribution, which describes the flux distribution for a PS population.\footnote{In the present discussion, we use flux, $F$, usually specified in units of [counts/cm$^2$/s], and counts, $S$, interchangeably.
The mapping between these quantities is controlled by the spatially-dependent instrument response, and while it must be done carefully, this is incorporated into the NPTF framework.
Explicitly, we do so using \texttt{NPTFit}, setting \texttt{nexp}=5.}
In the present work, we choose to parameterize the source-count distribution as follows:
\es{eq:mbpl}{
\frac{dN_p}{dF} ({\bm \theta}) = A\, T^{({\rm PS})}_p \left\{ \begin{array}{cc} 
\vspace{0.2cm}\left( \frac{F}{F_{b,1}} \right)^{-n_1} & F \geq F_{b,1} \\
\vspace{0.2cm}\left(\frac{F}{F_{b,1}}\right)^{-n_2} & F_{b,1} > F \geq F_{b,2} \\
0 & F_{b,2} > F  \\ 
\end{array} \right. \,.
}
Here, $T_p^{\rm (PS)}$ is the template that describes the overall expected spatial distribution of the sources.
For isotropically-distributed sources, the spatial template is constant over the sky ($T_p^{\rm (PS)} \propto 1$), while for spherically-distributed sources that may make up the GCE, the spatial template follows the observed morphology of the GCE.
The source-count distribution contains all the information about the population of sources; for example, we can use it to determine the expected number of sources in each pixel through $N_p = \int dF\,dN_p/dF \propto T_p^{\rm (PS)}$.

The parameters that determine the NPTF model in \eqref{eq:mbpl} are ${\bm \theta} = \{A, F_{b,1}, F_{b,2}, n_1, n_2\}$.
Importantly,  we do not allow PSs to have flux below $F_{b,2}$ in the chosen parameterization of the source-count distribution.  We stress that this is distinct from previous NPTF applications on data, \emph{e.g.}, Refs.~\cite{Lee:2015fea, Linden:2016rcf, Lisanti:2016jub, Leane:2019uhc}.  Our choice to remove the ultrafaint sources from the source-count distribution is motivated by the results of our companion paper~\cite{Chang:2019ars}.
In that work, we emphasized that there is a fundamental ambiguity between emission from a population of low-flux sources and smooth Poissonian emission.  This ambiguity becomes increasingly more pronounced as one approaches fluxes that correspond to single-photon sources; below this regime, emission from a population of unresolved PSs is simply Poissonian.  
In practice, this degeneracy introduces a fundamental ambiguity into questions regarding how much flux is associated with smooth emission and PSs with a similar spatial distribution due to potential biases in the NPTF parameterization.  Here, we conservatively decide to simply remove the low-flux end of the source-count distribution.  This means that any flux from PSs below $F_{b,2}$ would be absorbed by one or more Poissonian templates.
In practice, we do not treat $F_{b,2}$ as a free parameter and instead fix it at the approximate 1-$\sigma$ detection threshold for resolved PSs.

For all analyses, including those on both the real and simulated data, we use the likelihood described in Eq.~\eqref{eq:likelihood}.
For the Poissonian energy-binned analyses described in Sec.~\ref{sec:poiss}, we use a frequentist maximum-likelihood approach, where the maximum likelihood estimation is performed using \texttt{Minuit}~\cite{James:1975dr}.
For all NPTF analyses in Sec.~\ref{sec:harmNPTF} and \ref{sec:ROI}, we employ the likelihood in a Bayesian statistical framework, implemented using \texttt{MultiNest}~\cite{Feroz:2008xx,Buchner:2014nha}, setting the number of live points \texttt{nlive}=1000.
The priors on the templates and parameters are described in Sec.~\ref{sec:temps}.

\subsection{{\it Fermi} Dataset}

We make use of almost 8 years of data collected by the {\it Fermi} LAT.
The dataset consists of 413 weeks of the Pass 8 data, collected between August 4, 2008 and July 7, 2016.
We use the top quartile of \texttt{UltracleanVeto} data, as graded by the instrument PSF.  Note that we use the top quartile of data as opposed to including more quartiles because, while we would gain additional statistics by including more quartiles, \emph{(i)}~this would come at the expense of lower angular resolution, which may actually make it harder to find dim PSs, and \emph{(ii)}~we are already in the systematics-dominated regime (see~{\it e.g.}, App.~\ref{app:qual}).
The data is further subjected to the following conventional quality cuts: \texttt{DATA\_QUAL}==1, \texttt{LAT\_CONFIG}==1, and zenith angle $< 90^{\circ}$.
Finally, we only use photons with a reconstructed energy between 2 and 20 GeV.
This dataset is the exact one produced in Ref.~\cite{Mishra-Sharma:2016gis}, and it is publicly available as referenced there.
For several Poissonian analyses, we will use data separated into ten logarithmically-spaced energy bins over the same range. For the analyses that use PS templates, we work with only one single energy bin as it improves the ability to statistically distinguish the unresolved PSs.

Throughout this work, we do not use the full-sky data in our analyses.  Instead, we  restrict to specific regions of interest~(ROIs) that are relevant for studying the GCE, which only extends out from the Galactic Center~(GC) to $\mathcal{O}(10^{\circ})$~\cite{Daylan:2014rsa, Calore:2014xka}.
The fiducial region is defined by $|b|>2^{\circ}$ and $r < 25^{\circ}$, where $b$ is Galactic latitude and $r$ is the angle from the GC.  Note that this ROI overlaps closely with that used by Leane~and Slatyer~\cite{Leane:2019uhc}. 

In selecting the size of the ROI, one must carefully balance two separate concerns.  First, the ROI should overlap with the GCE region and be large enough to include enough photons to have statistical sensitivity to a population of unresolved PSs using the NPTF.  However, it should not extend too far beyond the GCE, because then the normalization of the diffuse foreground templates will be affected by data farther from the GC, making it more difficult for the foreground model to adjust to features near the GC. 
This issue is acute when using foreground models that are known to be imperfect, as is the case for any model of the {\it Fermi} diffuse emission.
This point has been discussed in the literature---for example in Refs.~\cite{Daylan:2016tia,Linden:2016rcf,Cohen:2016uyg,Lisanti:2017qlb,Lisanti:2017qoz,Chang:2018bpt}.
A large focus of this paper is to demonstrate how the NPTF results vary as a function of the ROI.  We achieve this in two ways.  The first is by giving the diffuse models additional degrees of freedom to adjust to large-scale variations, using a spherical-harmonic analysis (Sec.~\ref{sec:harm}).  The second is a simpler test where we explore the impact of ROI size by varying the cut on $r$ (Sec.~\ref{sec:ROI}).  

In addition to restricting our analyses to a specific spatial region, we also excise known (\emph{i.e.}, resolved) PSs from the ROI.
To do so, we take all sources detected in the {\it Fermi} 3FGL catalog~\cite{Acero:2015hja} and mask a ring around the location of each PS that corresponds to the 95\% containment radius of the instrument PSF for our dataset at 2~GeV.
The {\it Fermi} Collaboration recently released an update to the 3FGL catalog, the 4FGL~\cite{Fermi-LAT:2019yla}, and we explore the impact on our results of masking the PSs in this catalog in App.~\ref{app:4FGL}.  We do  not mask the 4FGL catalog in the fiducial analysis because doing so removes a significant fraction of the available ROI in the inner few degrees.  Along similar lines, we also note that Ref.~\cite{Lee:2015fea} masked the 3FGL sources at a much larger containment fraction than we do here.  Consequently, for a given 3FGL source, the resulting contribution to the PS mask was over four times as large in Ref.~\cite{Lee:2015fea} relative to this work.  We chose the present masking scheme because the 95\% containment radius is sufficient to mask the 3FGL sources while maintaining sufficient area in the inner regions of the Galaxy.  Because of this, certain details regarding the 3FGL-masked analyses presented here (such as \emph{e.g.}, Bayes factors) are not directly comparable to those in Ref.~\cite{Lee:2015fea}.    

All analyses that we perform on {\it Fermi} data are calibrated on simulated datasets, before being applied to the real data.
It is straightforward to simulate gamma rays from pure Poissonian emission: we simply take a pixel-by-pixel Poisson draw from the sum of the relevant model templates.
For emission associated with any PS template, the procedure is more involved, and throughout we generate the simulated data using the code package \texttt{NPTFit-Sim}.\footnote{Publicly available  \href{https://github.com/nickrodd/NPTFit-Sim}{here}.}

\subsection{Spatial Templates and Model Priors}
\label{sec:temps}

The central input of any template analysis is the assumed spatial distribution of each emission component.
As discussed above, the model parameters do not incorporate any freedom for varying the shape of each  template.
Consequently, deviations between the assumed maps and the true distributions are fundamental systematics for a template analysis.
In the next subsection, we will introduce a novel technique for addressing this source of uncertainty.
Our goal for now is to introduce the fiducial templates, paying particular attention to the collection of models we consider for the diffuse foreground.

For the NPTF analyses, the priors adopted on the models described in this section are shown in Tab.~\ref{tab:priors}.
The predicted emission also depends on how the spatial templates $T_p$  (whether Poissonian or non-Poissonian) are normalized.
Following Ref.~\cite{Mishra-Sharma:2016gis}, we take all templates to have a mean value of unity (in terms of counts) in the ROI defined by $|b|>2^{\circ}$ and $r < 30^{\circ}$. %
We note, however, these the template normalizations are arbitrary and do not affect the final results (in detail, they are only of relevance for interpreting our prior choices).

Beyond the Galactic diffuse emission templates, described below, we use four Poissonian templates.
The first of these is an isotropic template, motivated by extragalactic emission that is expected to be roughly uniform across the sky.
The second is a template to capture the emission associated with the {\it Fermi} bubbles~\cite{Su:2010qj}.
The third is a Poissonian template for the known 3FGL PSs.  Even though these sources are masked up to their 95\% containment radius, this template captures any emission that may extend beyond these masks---this is especially relevant for the brightest sources.
Finally, as we are studying the DM interpretation of the GCE, we include a template that has the spatial profile expected for annihilating DM.
In detail, this template traces the square of a generalized NFW profile, integrated along the line of sight.
The generalized NFW profile density is given by
\be
\rho(r) \propto \frac{1}{(r/r_s)^{\gamma}(1+r/r_s)^{3-\gamma}}\,,
\ee
where the scale radius is $r_s = 20$~kpc.
The canonical NFW profile has $\gamma=1$, but motivated by previous studies~\cite{Daylan:2014rsa,Linden:2016rcf,Keeley:2017fbz}, we adopt a profile with $\gamma=1.2$.
The choice of $\gamma=1.0$ versus $1.2$ does not qualitatively affect our primary conclusions.

As indicated in Tab.~\ref{tab:priors}, we generically allow the normalization of the Poissonian GCE template to go negative.  As discussed later in this work, negative normalizations for the GCE template can be symptoms of over-subtraction induced by diffuse mismodeling.  With that said, when we compute Bayes factors comparing models with and without GCE-correlated PSs, we restrict the Poissonian GCE normalization to be strictly positive.  The reason is that when determining whether one model fits the data better than another, it is better  justified to restrict the models to their physical parameter spaces, and physically the GCE should be positive if {\it e.g.}, it arises from DM annihilation.

\begin{table}[tb]
\footnotesize
\begin{center}
\begin{tabular}{C{2cm}C{2cm} | C{2cm}C{2cm}}
\multicolumn{2}{c}{\textbf{Poissonian}} & \multicolumn{2}{c}{\textbf{Non-Poissonian}}  \Tstrut\Bstrut	\\   
\textbf{Parameter}	 & \textbf{Prior}  & \textbf{Parameter}	&  \textbf{Prior}   \Tstrut\Bstrut	\\   
\Xhline{1\arrayrulewidth}
$A_{{\rm dif}/\pi^0}$  & [5,\,30]  & $\log_{10} A^{\rm (PS)}$ & [-6,\,1] \Tstrut\Bstrut \\
$A_{\rm ics}$ & [0,\,15]  &  $n_1$ & [2.05,\,10]  \Tstrut\Bstrut \\ 
$A_\text{iso}$  & [-30,\,30]  & $n_2$ & [0.05, 3.5] \Tstrut\Bstrut \\ 
$A_\text{bub}$& [0,\,2] &  $S_{b,1}^{\rm GCE}$ & [$S_{b,2}$, 40] \Tstrut\Bstrut \\
$A_\text{GCE}$ & [-5,5] &  $S_{b,1}^{\rm Disk}$ & [$S_{b,2}$, 60]     \Tstrut\Bstrut \\
$A_\text{3FGL}$ & [0,\,10] &   & \Tstrut\Bstrut \\
\end{tabular}
\end{center}
\caption{Priors adopted on the templates described in Sec.~\ref{sec:temps}.
For the non-Poissonian templates, we fix the lower-flux cutoff for the source-count function ($S_{b,2}$) to the approximate 1-$\sigma$~PS detection threshold, and allow no sources with an expected count below this. This is motivated by the results of our companion paper~\cite{Chang:2019ars}, which highlighted the fact that PSs become degenerate with smooth DM emission in the ultrafaint limit.  As implemented in this paper, all emission from PSs with flux below $S_{b,2}$ would look like pure Poissonian emission to the NPTF. }
\label{tab:priors}
\end{table}

In addition to the Poissonian templates described above, we use two non-Poissonian templates.  The first template describes the possibility that the GCE arises from PSs as opposed to smooth emission, \emph{e.g.}, due to DM annihilation.  That template is the same as described above for the Poissonian DM annihilation model.
We also employ a second non-Poissonian model to trace a PS population that is correlated with the disk of the Milky Way.
For this purpose, we assume the disk sources follow a doubly-exponential profile:
\be
n(R,z) \propto \exp (-R/5~{\rm kpc})\,\exp (-|z|/1~{\rm kpc})\,,
\ee
where $R$ and $z$ are the radial and vertical Galactic cylindrical coordinates, and then by integrating this profile along the line of sight, we construct an appropriate $T_p^{\rm (PS)}$. This spatial profile is motivated by studies of the disk distribution of millisecond pulsars~\cite{Lorimer:2006qs, Bartels:2018xom}.   Note that, we do not include the 3FGL Poissonian template  for the NPTF studies, as we allow the additional PS flux to be absorbed by the disk PS template.

Finally, we turn to the sky maps used for the Poissonian models associated with the Galactic diffuse emission.
The Galactic diffuse emission accounts for the bulk of the photons detected by the {\it Fermi} LAT, particularly towards the GC.
As described above, there are three primary mechanisms accounting for this emission.
The dominant contribution arises from the interaction of cosmic-ray protons with gas, producing pions, the neutral variants of which then decay to photons.
A second source is the bremsstrahlung emission resulting from the interaction of cosmic-ray electrons with the same gas.
Both of these contributions to the Galactic emission are largely correlated with tracers of the interstellar gas.  The interstellar gas is dominated by atomic hydrogen (H$_{\rm I}$), which is traced by 21-cm line emission, and molecular hydrogen (H$_{\rm II}$), which is traced by the 2.6~mm line emission from carbon monoxide (CO).
The third source for the emission is associated with the same electron population up-scattering off the cosmic microwave background and interstellar radiation fields of the Milky Way via IC scattering, and the sky map associated with this component is controlled by the distribution of the electrons and radiation fields.
Given the above contributors, the diffuse models we consider will either be described by a single template incorporating all three contributions, or alternatively two separate templates, one correlated with the gas accounting for the $\pi^0$ and bremsstrahlung emission, and a second map describing emission due to the IC process.

Given that the diffuse emission is the dominant source of systematic uncertainty, we consider several different models in this work.  
The first is one of the official {\it Fermi} diffuse models: \texttt{gll\_iem\_v02\_P6\_V11\_DIFFUSE} (\texttt{p6v11}).\footnote{\href{https://fermi.gsfc.nasa.gov/ssc/data/access/lat/ring_for_FSSC_final4.pdf}{https://fermi.gsfc.nasa.gov/ssc/data/access/lat/ring\_for\_FSSC\\ \_final4.pdf}}  The \texttt{p6v11} model is built by fitting empirical H$_{\rm I}$ and CO maps to the {\it Fermi} data in six Galactocentric rings, while modeling the IC component using \texttt{GALPROP}~\cite{Strong:1998fr,Porter:2017vaa}.\footnote{\href{https://galprop.stanford.edu}{https://galprop.stanford.edu}}  For \texttt{p6v11}, the individual gas and IC-correlated components are not provided and so we cannot vary over their individual normalizations in the analysis.  Note that \texttt{p6v11} is one of the most common diffuse models used in GCE studies, including the original NPTF paper~\cite{Lee:2015fea}, and was the primary model used by Leane and~Slatyer in Ref.~\cite{Leane:2019uhc}.  
The reason for this is that \texttt{p6v11} is the last official {\it Fermi} diffuse model that does not include  large-scale structures like the {\it Fermi} bubbles.  A particular concern is that such large-scale structures, when determined in a data-driven manner, may have overlap with the GCE.  In general, one should avoid situations where the GCE may be accidentally incorporated into the diffuse model.  However, concerns regarding \texttt{p6v11} have already been discussed in the literature; for example Ref.~\cite{Calore:2014xka} noted that \texttt{p6v11}'s hard IC component above 10~GeV can lead to potential over-subtraction in the data, shaping the high-energy tail of the GCE.

In addition to \texttt{p6v11}, we will make use of Model~A and F, which were used in Ref.~\cite{Calore:2014xka}, although Model~F was originally generated in Ref.~\cite{Ackermann:2012pya}.
In these two cases, both the gas-correlated and IC emission are generated with \texttt{GALPROP}.  There are a variety of assumptions that must be specified when modeling the emission components in this way, including information on the source and gas distribution, the interstellar radiation field, the magnetic field distribution, and parameters associated with diffusion, re-acceleration, and convection.  The specific parameter choices for Models A~and~F are detailed in Ref.~\cite{Calore:2014xka}.   
As was shown in that work, compared to \texttt{p6v11}, Model~A is a better fit to the {\it Fermi} data at energies above 1~GeV, while Model~F is a better fit at all energies.

\begin{figure*}[htb]
\leavevmode
\begin{center}
\includegraphics[width=.47\textwidth]{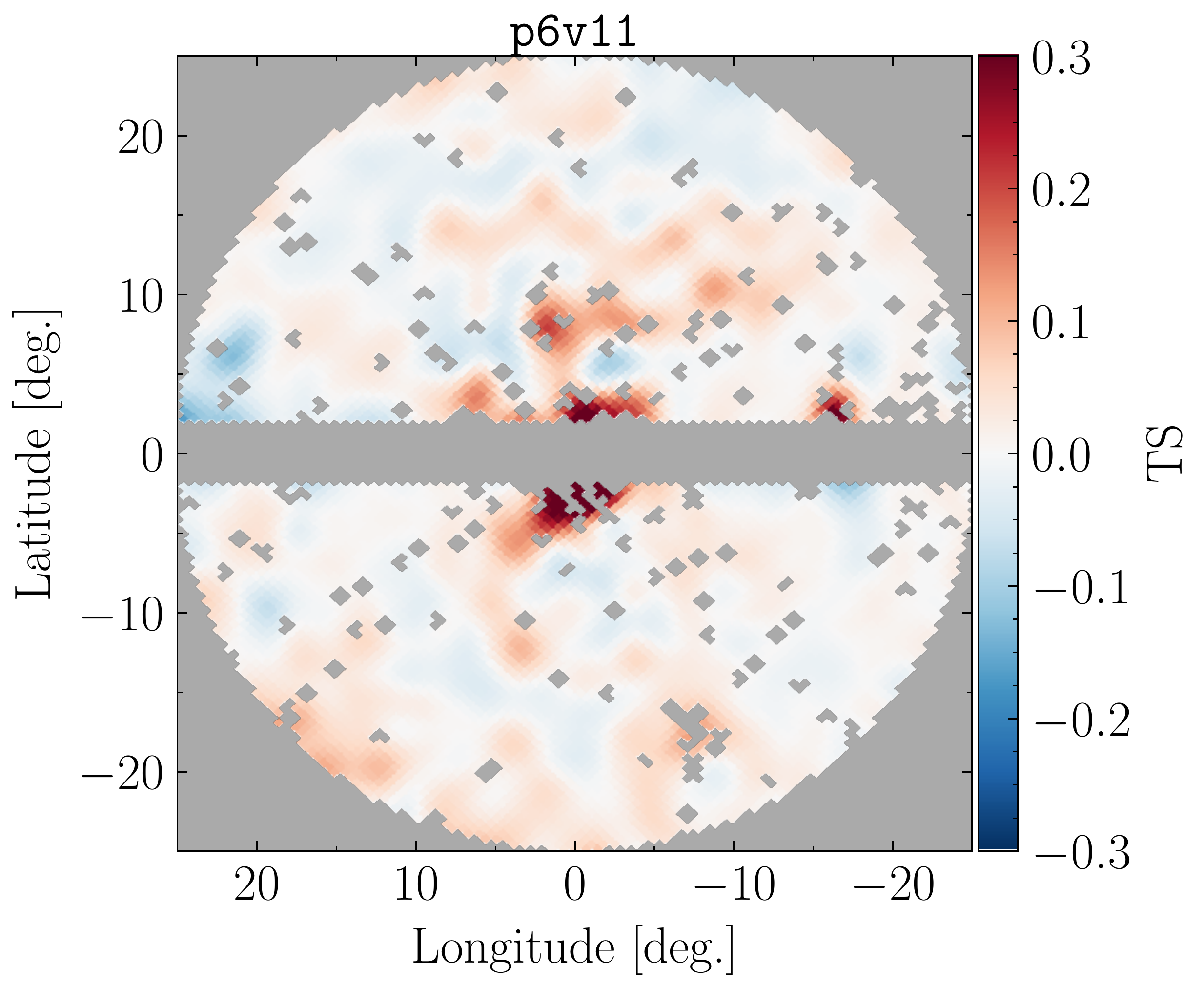} \hspace{0.5cm}
\includegraphics[width=.47\textwidth]{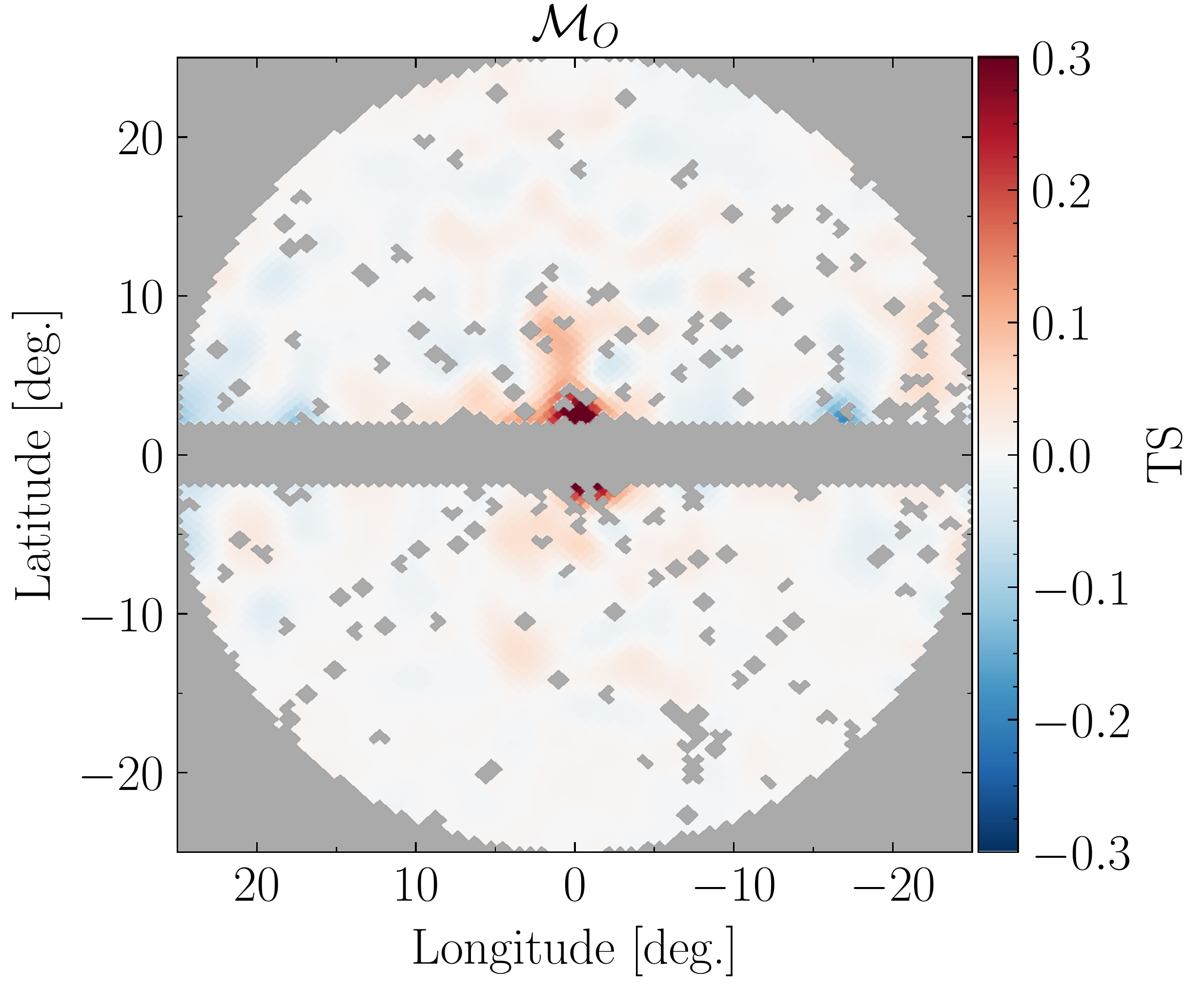}
\end{center}
\caption{A visualization of where spatially the preference for a non-Poissonian GCE template draws its power for two different diffuse models: \texttt{p6v11} (left) and Model~O, $\mathcal{M}_O$~(right).  See text for details on the differences between these two model scenarios. 
In detail, each map shows the pixel-wise TS, defined as twice the log-likelihood difference between an analysis with and without a non-Poissonian GCE template.
Gray pixels are masked and not included in the analysis.
The final map is smoothed using a $1^{\circ}$ Gaussian that ignores masked pixels.
There are two primary conclusions that can be drawn from these figures.  The first is that the statistical preference for an unresolved population of GCE PSs is driven by the inner $\sim$$5^\circ$ of the Galaxy.  The second is that in order to accommodate the GCE PS template, the \texttt{p6v11} model introduces more large-scale restructuring compared to Model~O.
}
\label{fig:TS-Maps}
\end{figure*}

The final diffuse model we consider is Model O, which we construct ourselves.\footnote{Here, we are breaking with the naming convention of~\cite{Calore:2014xka}, where a diffuse model referred to as Model~O was considered, but is different than what we use here.}
Model~O consists of a linear combination of templates representing components of the Galactic diffuse emission.
In particular, interstellar gas-correlated photons are modeled using templates of H$_{\rm I}$ and H$_{\rm II}$ obtained from a suite of hydrodynamical simulations of interstellar gas material~\cite{Pohl:2008}.
The construction of such maps relies on two simplifying assumptions: \emph{(i)}~that molecular hydrogen is well-mixed with CO; and \emph{(ii)}~that the atomic hydrogen spin temperature $T_{S}$ is constant throughout the Galaxy.
Both assumptions are expected to affect the estimates of interstellar gas column density since the spin temperature can vary along a certain line of sight and the spatial correlation of CO with H$_{\rm II}$ may not be perfect.
In order to correct for these deficiencies, we also consider dust residual (or dark gas) templates constructed using methods introduced in Ref.~\cite{Ackermann:2012pya}.
Refs.~\cite{Macias:2016nev,Macias:2019omb} showed that there are morphological differences between the hydrodynamic gas maps and the standard gas maps included in the {\it Fermi} diffuse emission model, and that the former are statistically preferred by gamma-ray data from the GC region.
The IC template associated with Model~O is constructed using the most recent 3D interstellar radiation field models in {\tt GALPROP} v56~\cite{Porter:2017vaa}.
We choose the Galaxy-wide dust and stellar distribution model based on Ref.~\cite{Freudenreich:1997bx} and the intermediate cosmic-ray propagation setup called SA50 (see Table 3 of Ref.~\cite{Porter:2017vaa}).

Once constructed, both the gas and IC templates are subdivided into Galactocentric rings (assuming radial ranges of 0--3.5, 3.5--8, 8--10 and 10--50~kpc).
These are included separately in the fitting procedure to account for cosmic-ray density variations with distance and reduce the impact on the results of the choice of {\tt GALPROP} propagation parameter setup.
When performing energy-dependent Poissonian scans, we give Model~O fourteen degrees of freedom.
In detail, twelve of these arise as we allow the four Galactocentric rings associated with the H$_{\rm I}$, H$_{\rm II}$, and IC maps to float separately.
The final two degrees of freedom are associated with the dust residual templates, which are not divided into rings.
For these runs we use \texttt{Minuit} to find the maximum likelihood, and as this is a frequentist procedure we do not need to specify the priors on each template.
However, for our non-Poissonian analyses, we combine the rings according to their best-fit values in the Poissonian scan, as determined by the maximum likelihood, in each energy bin.
We combine the eight H$_{\rm I}$ and H$_{\rm II}$ maps, together with the two dust residual maps, to form a combined gas-correlated template designed to model the $\pi^0$ and bremsstrahlung emission.
We combine the four IC rings separately to make up an IC template.
As such, for the non-Poissonian analyses Model~O has two degrees of freedom, similar to Model~A and F.\footnote{The final $\pi^0$ and IC Model O templates we use in our default ROI are available \href{https://github.com/nickrodd/FermiDiffuse-ModelO}{here}.}

As we will see in the next section, of the four models for the diffuse emission considered, \texttt{p6v11} provides by far the worst fit to the data.
Nevertheless, it is an important model to consider, as it is used in many of the canonical GCE studies.
A central theme of the present work is how the four diffuse models introduced above perform across various benchmarks. To start, we highlight one way of visualizing the improvements provided by, for example, Model~O over \texttt{p6v11} in Fig.~\ref{fig:TS-Maps}.
This figure helps to visualize where the NPTF draws its power from across the sky.
Both maps show the pixel-wise test statistic (TS), defined as twice the log-likelihood ratio between models with and without a GCE non-Poissonian template.  In both cases, we also include all best-fit base Poissonian templates (including a Poissonian GCE model) and a non-Poissonian disk template.
The resulting TS map is then smoothed using a Gaussian of $1^{\circ}$ width, ignoring masked pixels, and we show the result for two diffuse models, \texttt{p6v11} and Model~O.  The NPTF is performed over the full canonical ROI
and the TS maps are then computed at the medians of the posteriors for the model components.  Larger values for the TS indicate that including the GCE non-Poissonian template improves the goodness-of-fit at that spatial location, while negative values imply that the fit is worsened at that location by the inclusion of the additional non-Poissonian model.

\begin{figure*}[htb]
\leavevmode
\begin{center}
\includegraphics[width=.97\textwidth]{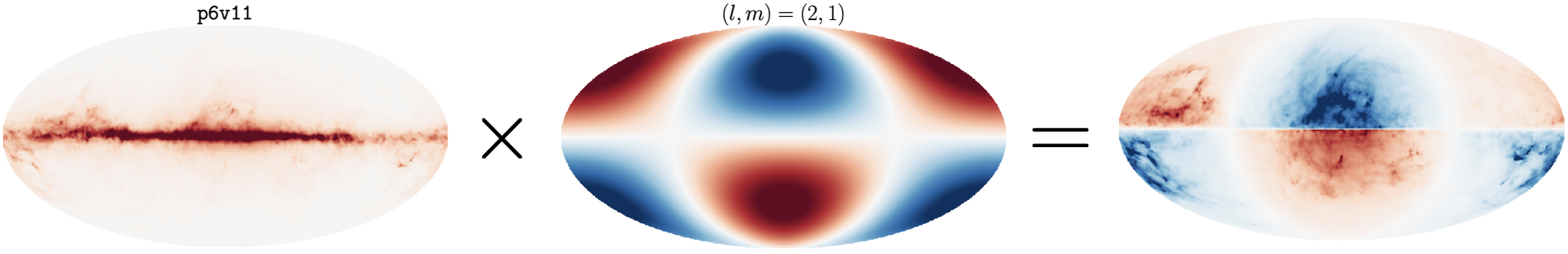}
\end{center}
\caption{Depiction of how the harmonic diffuse maps are constructed.
For the example of \texttt{p6v11} (left), we combine the map with the $l=2$, $m=1$ spherical-harmonic map (middle), to produce the hybrid harmonic diffuse sky map (right).}
\label{fig:Harm}
\end{figure*}

There are two immediate conclusions that can be drawn from Fig.~\ref{fig:TS-Maps}.
First, is that the evidence for a PS origin of the GCE is strongly driven by the inner $\sim$5$^\circ$ around the GC.
This is the region where the diffuse emission is expected to be the most uncertain, and thus raises the stakes for minimizing the impact of this systematic. 
A new method for doing exactly that is introduced in the next subsection.  Second, we see that when using the \texttt{p6v11} diffuse model, including the GCE PSs leads to a large-scale restructuring of the emission throughout the ROI.
As we show in the following sections, although \texttt{p6v11} provides greater evidence for the GCE non-Poissonian model than the other diffuse models, this evidence is partly an artifact of the large-scale mismodeling in \texttt{p6v11}.

\subsection{Harmonic Marginalization}
\label{sec:harm}

In this paper, we present a new method to account for large-scale mismodeling of diffuse emission templates in a data-driven fashion.
The basic idea is that for PS searches, we can marginalize over uncertainties at larger angular scales without affecting our ability to find the small-scale structures of interest.
Large-scale mismodeling of {\it e.g.}, the diffuse foreground may affect our ability to find PSs  because when large-scale mismodeling is present then the diffuse model will both over- and under-predict the data at various locations.

There are multiple ways in which the diffuse model may be given more degrees of freedom to account for large-scale uncertainties.
In Ref.~\cite{Cohen:2016uyg}, the diffuse emission was given independent degrees of freedom above and below the Galactic plane, leading to a significantly improved fit.
In Ref.~\cite{Chang:2018bpt}, the diffuse model was divided into independent spatial regions and each component was given its own nuisance parameter.
Refs.~\cite{Storm:2017arh, Bartels:2017vsx} included a large number of nuisance parameters to allow spatial and spectral modulation of the diffuse emission, using regularization techniques to impose physicality conditions.
In this work, we consider an alternate method that accomplishes the same goal.
We construct a sequence of spatial templates by multiplying the original diffuse model (or any other Poissonian template that may suffer from large-scale mismodeling effects) $T^{\rm diff}(\theta, \phi)$ by spherical harmonics $Y^{\ell,m}(\theta,\phi)$ to construct the set of templates.
Of course, as both maps are pixelized, the combined template is $Y^{\ell,m}_p T^{\rm diff}_p$.
An example of a template constructed in this manner is shown in Fig.~\ref{fig:Harm}.  In this case, the \texttt{p6v11} template~(left panel) is multiplied by the $l=2, m=1$ spherical-harmonic map~(middle panel) to yield the final template~(right panel) used in the analysis.

Each harmonic template map is assigned its own nuisance parameter $A_{\ell,m}$, corresponding to the normalization of these maps.
We only consider templates up to some maximum $(\ell_{\rm max}, m_{\rm max})$ in order to marginalize over uncertainties at large angular scales.
We marginalize over the $A_{\ell,m}$ when constraining the physical model parameters of interest; the detailed procedure is described below.
In Sec.~\ref{sec:poiss}, we show how this method allows for a more consistent determination of the GCE spectra between diffuse models in a purely Poissonian analysis, and then in Sec.~\ref{sec:harmNPTF}, we apply this method to the NPTF and show that it gives a consistent PS interpretation of the GCE amongst diffuse models considered.   
For larger values of $\ell_{\rm max}$ and $m_{\rm max}$, the number of harmonic templates can become considerable.
In each instance, we perform an initial purely Poissonian run using \texttt{Minuit}.
From this fit, we extract the template normalizations that achieve the maximum likelihood, denoted $\hat{A}_{\rm diff}$ and $\hat{A}_{\ell,m}$.
From these, a single harmonically improved template is formed as follows:
\be
T_p^{\rm harm} \propto \hat{A}_{\rm diff} T^{\rm diff}_p + \sum_{\ell,m} \hat{A}_{\ell,m} Y^{\ell,m}_p T^{\rm diff}_p\,,
\ee
which we can then normalize as desired.
This single improved map is then what we use in the non-Poissonian run.

When performing the harmonic marginalization, we envision these corrections as being relatively small corrections to the diffuse modeling rather than $\mathcal{O}(1)$ corrections.
To ensure this, we add a Gaussian penalty (regularization) term to the likelihood.
In detail, for each harmonic template we multiply the likelihood by
\be
\mathcal{L}_{\rm penalty} = \frac{1}{\sigma \sqrt{2\pi}} \exp \left[ - \frac{A_{\ell,m}^2}{2\sigma^2} \right]\,,
\ee
where we take $\sigma$ to be 20\% of the best-fit \texttt{p6v11} diffuse model normalization in the case without harmonics.
Note that we are biasing the fit to prefer $A_{\ell,m}=0$, as the spherical harmonics are both positive and negative across the sky.

\begin{figure*}[htb]
\leavevmode
\begin{center}
\includegraphics[width=.49\textwidth]{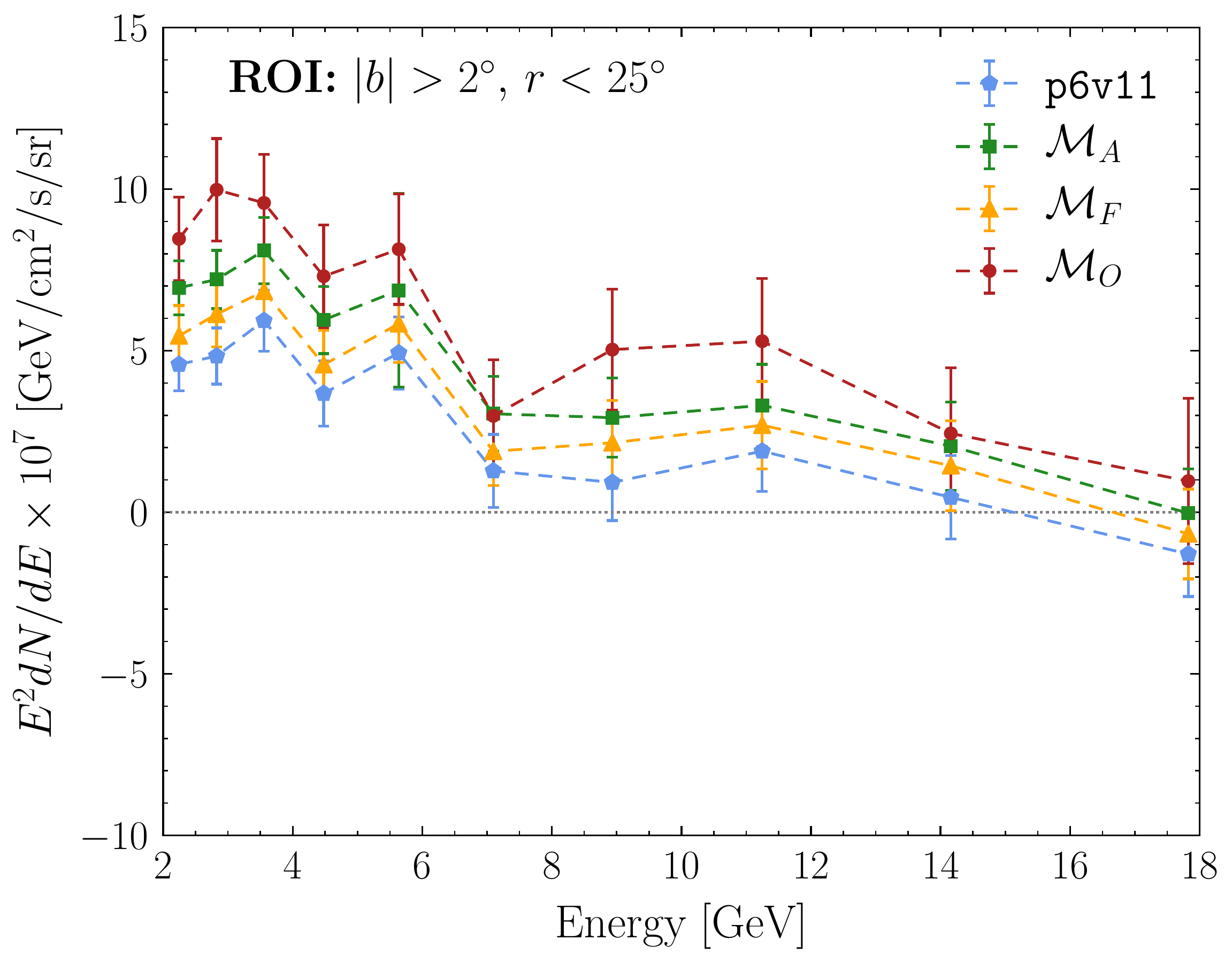} \includegraphics[width=.49\textwidth]{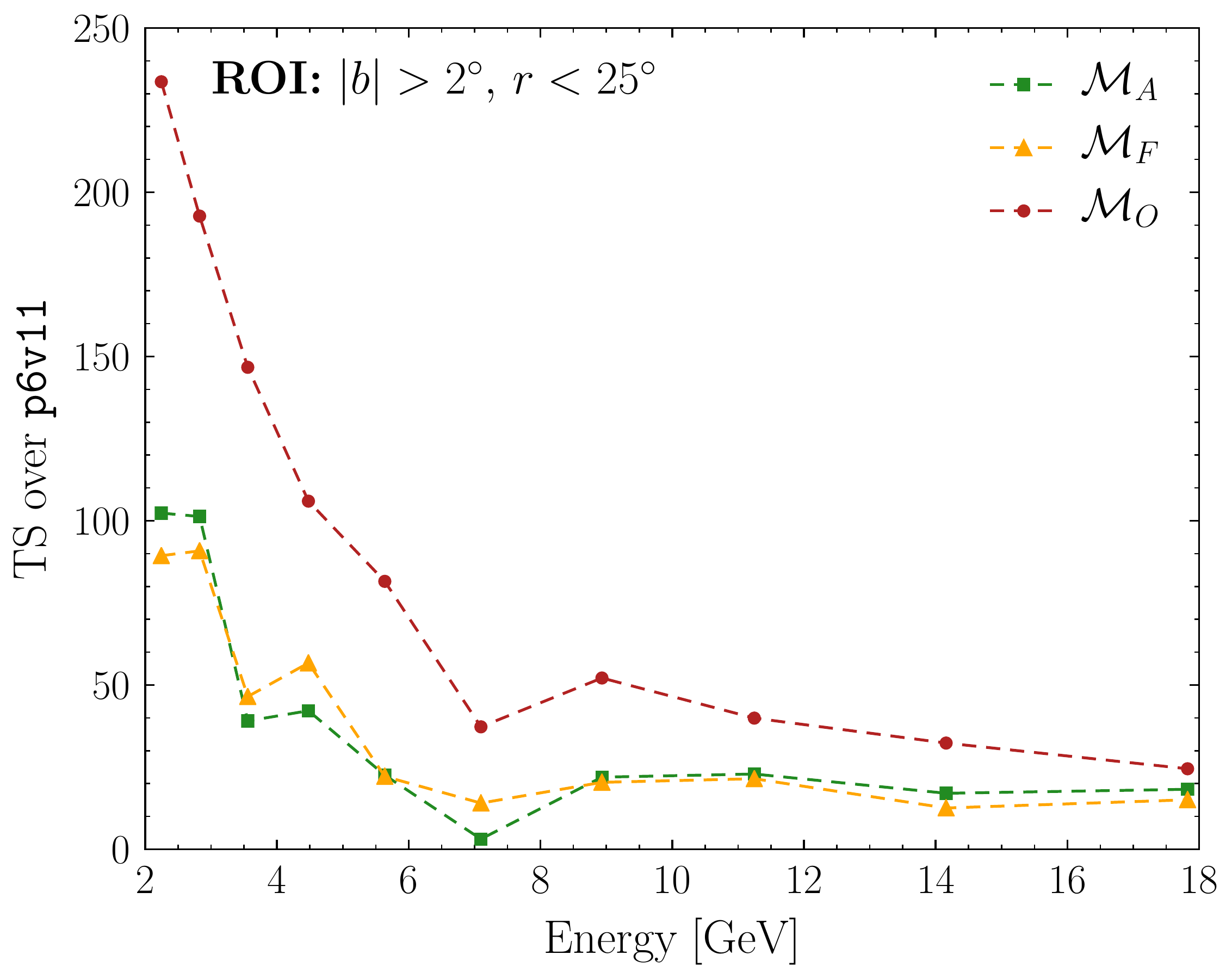}
\end{center}
\caption{(Left) Spectrum of the average emission associated with the Poissonian GCE extracted as a function of energy within our fiducial ROI ($|b| > 2^\circ$, $r< 25^\circ$) for the four different diffuse models studied: \texttt{p6v11}, as well as Models A ($\mathcal{M}_A$), F ($\mathcal{M}_F$), and O ($\mathcal{M}_O$).  These are designated by the dashed lines in blue, green, orange, and red, respectively.  The analyses performed here are purely Poissonian, and include templates for diffuse emission, isotropic emission, the {\it Fermi} bubbles, 3FGL PSs, and a fiducial GCE template (modeled assuming an NFW profile).  We find evidence for the GCE across all diffuse models, though the normalization can vary by as much as a factor of $\sim 2$ between them and is highest for Model~O.  As already underscored  in Ref.~\cite{Calore:2014xka}, care must be taken when interpreting the GCE because the systematic uncertainties from modeling the diffuse emission are greater than the statistical uncertainties, indicated by the error bars.  (Right) The TS in favor of a given diffuse model over  \texttt{p6v11}.  The TS is computed by comparing the log-likelihoods at the best-fit points from the fits that go into the left panel.  Models A, F, and~O outperform \texttt{p6v11} across all energy bins above 2~GeV, and Model~O provides the best fit to the data.
}
\label{fig:GCE-spec}
\end{figure*}

\section{Poissonian analysis of the GCE}
\label{sec:poiss}

In this section, we show that properties of the GCE, as recovered from a purely Poissonian template analysis, are strongly affected by the choice of diffuse model and ROI.  In particular, we show that certain diffuse models suffer from over-subtraction similar to what was observed by Leane~and~Slatyer~\cite{Leane:2019uhc}, but for the purely Poissonian case.  We then apply the harmonic marginalization procedure described in the previous section and demonstrate that these specific over-subtraction issues are resolved.
 
Spectral and morphological studies of the dependence of the GCE on diffuse models have been carried out before, such as in the dedicated study in Ref.~\cite{Calore:2014xka}.  However, our focus here is to establish a few specific points that go beyond these earlier works.  One point is simply that diffuse models are now available that provide a significantly better fit to the data in the Inner Galaxy than the \texttt{p6v11} model, and that the evidence for the GCE is robust even with these newer models.  The second point is that diffuse mismodeling can lead to over-subtraction in the Poissonian template analyses.  We show explicitly that the \texttt{p6v11} diffuse model in particular suffers from over-subtraction in the outer region of the Inner Galaxy, whereby the GCE template prefers large negative values.  However, the harmonic marginalization procedure is able to mitigate the over-subtraction issue for \texttt{p6v11}.

\subsection{GCE Spectrum for Varying Diffuse Models} 
\label{sec:Poiss-fid}

To begin, we perform a standard Poissonian template analysis to recover the GCE energy spectrum in ten log- spaced bins from 2--20~GeV using the four benchmark diffuse models: \texttt{p6v11} and Models~A, F, and O.  We restrict ourselves to the fiducial ROI ($r \leq 25^\circ$, $|b| \geq 2^\circ$, with 3FGL PSs masked).
In addition to the templates associated with the diffuse emission, we also include templates for isotropic emission, the {\it Fermi} bubbles, and 3FGL PSs (to absorb any emission beyond the PS mask).  Additionally, we include the fiducial GCE template, modeled using the NFW DM profile previously discussed.  

The left panel of Fig.~\ref{fig:GCE-spec} shows the energy spectra that we recover, normalized with respect to the fiducial ROI.  Consistent with previous studies such as Ref.~\cite{Calore:2014xka}, we see that while the normalization of the GCE depends on the diffuse model used in the analysis, it is always non-zero between $\sim$2--8~GeV, within statistical uncertainties.  However, the normalization of the GCE can vary by as much as a factor of two between the models we explore.  In particular, Model~O has the highest normalization, while \texttt{p6v11} has the lowest.  This variation between models is perhaps not too surprising when considering that the diffuse foregrounds make up the vast majority of photon emission within the ROI.  Still, this result underlines that care must be taken when interpreting the GCE, considering that systematic uncertainties from diffuse mismodeling are far greater than the statistical uncertainties.   

The right panel of Fig.~\ref{fig:GCE-spec} illustrates which diffuse model provides a better fit to the data in the fiducial ROI.  This figure shows the TS in preference for a specific diffuse model compared to \texttt{p6v11}.  The TS is evaluated by comparing twice the log-likelihood associated with the best-fit point in a given energy bin when the analysis is run using Models~A, F, or O as opposed to  \texttt{p6v11}.
Model~O provides by far the best fit to the data over all energies above 2~GeV.  Models~A and~F also fit the data significantly better than \texttt{p6v11} in this ROI.  Model O has 14 degrees of freedom per energy bin whereas naively \texttt{p6v11} has one, so even in the most straightforward way of interpreting the change in the TS per degree of freedom we see that Model O is a better fit to the data (the change in the TS per degree of freedom in going to Model O is greater than unity).  However, the \texttt{p6v11} model was constructed itself from fits to the {\it Fermi} data with additional degrees of freedom, so even that counting (giving Model~O 13 more degrees of freedom than \text{p6v11}) is likely overly biased against Model O.

\subsection{Over-subtraction of the GCE}
\label{sec:Poiss-outer}

Mismodeling the diffuse emission can significantly affect the spectrum of the GCE, as we have already seen, and can also lead to what is called over/under-subtraction.  Over-subtraction occurs when a given template (in this case, the GCE template) is driven to lower-than-physical normalization due to mismodeling of other emission components (in this case, the diffuse model).  This arises because the mismodeled template erroneously absorbs more flux than it should.  Under-subtraction is the related effect whereby the GCE template has a larger-than-physical normalization because the diffuse template absorbs too little flux.  As all the spectra are positive in Fig.~\ref{fig:GCE-spec}, we cannot say definitively if any of the diffuse templates suffer from these issues because we do not know the true spectrum of the excess.  

\begin{figure}[t]
\leavevmode
\begin{center}
\includegraphics[width=.49\textwidth]{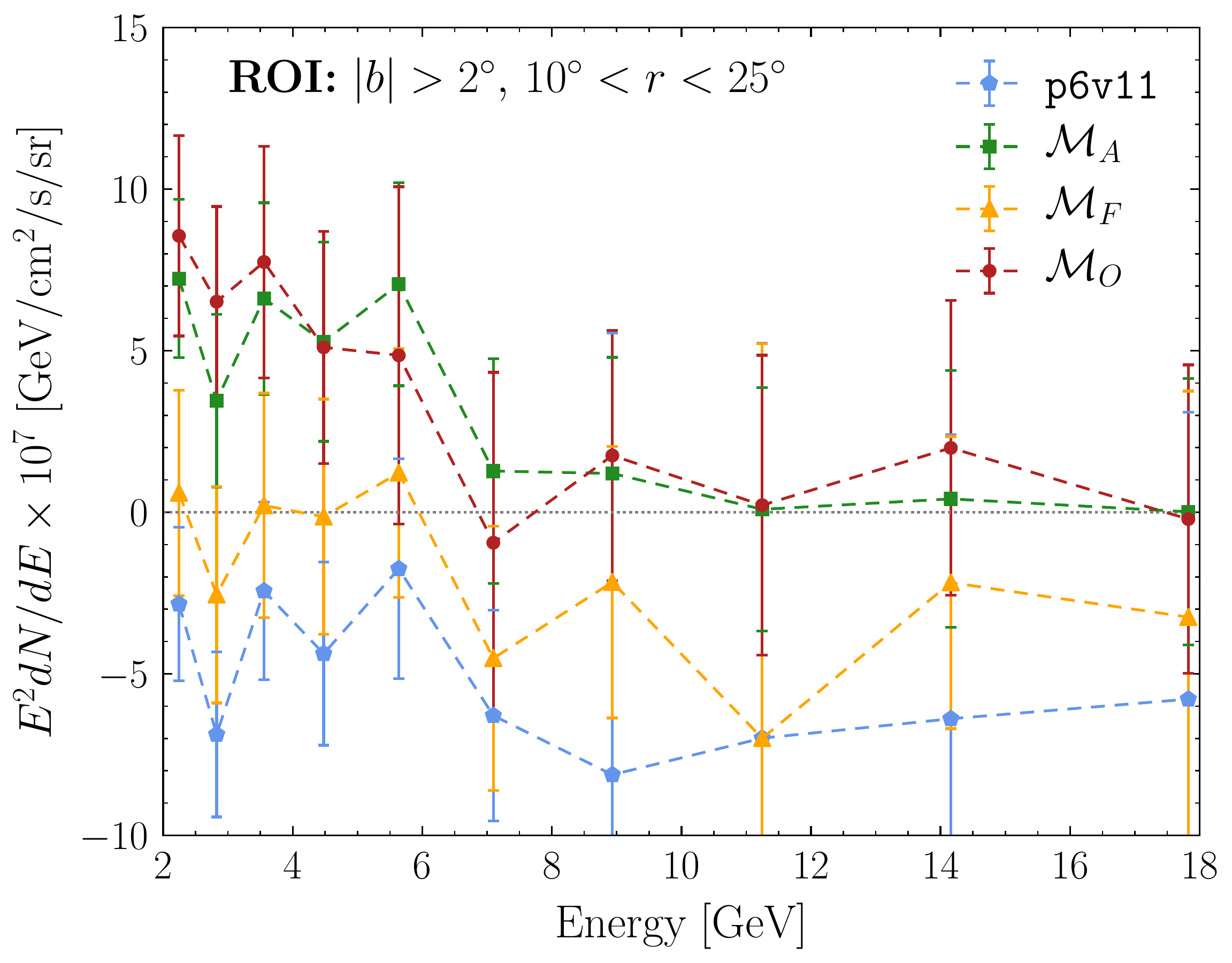} 
\end{center}
\caption{
As in the left panel of Fig.~\ref{fig:GCE-spec}, but for the restricted ROI with the inner 10$^\circ$ masked.  Note, however, that the fluxes are still computed relative to the fiducial ROI so that these spectra can be easily compared to those in the left panel of Fig.~\ref{fig:GCE-spec}.  One expects that the spectra should be consistent between the two ROIs if the foregrounds are well-modeled.  We see that this is true for Models~A and~O, within uncertainties.  However, \texttt{p6v11} and (to a lesser extent) Model~F clearly suffer from over-subtraction.  This is apparent from the fact that the recovered fluxes for the GCE are negative for most energies (indeed all for \texttt{p6v11}), indicating that the diffuse model template has absorbed too much flux and driven the GCE template to unphysical values.
}
\label{fig:GCE-spec-outer}
\end{figure}

However, we do know for certain that the GCE spectrum must be positive or consistent with zero flux.  If the GCE spectrum is driven to significantly negative values for certain diffuse models, then that is a clear indication that those diffuse models suffer from over-subtraction.  As it turns out, both \texttt{p6v11} and Model~F do suffer from over-subtraction in the outer regions of the fiducial ROI.  To illustrate this point, we repeat the analyses presented in Sec.~\ref{sec:Poiss-fid}, but also requiring that $r \geq 10^\circ$.  The results of this analysis are shown in Fig.~\ref{fig:GCE-spec-outer}.  Note that even though we perform the analyses in the restricted ROI, we normalize the spectrum $E^2 dN/dE$ to the fiducial ROI to facilitate a comparison with the left of Fig.~\ref{fig:GCE-spec}.

If the diffuse models describe the data at the level of Poisson noise, then we expect the spectra to be consistent between the fiducial and restricted ROI analyses.  While this is true for Models~A and~O, it is certainly not the case for \texttt{p6v11} and, to a lesser extent, for Model F.  In the latter two cases, the energy spectrum of the GCE is consistently negative across all energy bins. Since the energy spectrum of the GCE can physically not be negative, this indicates that the diffuse models in these cases are not a good description of the data in the ROI and are systematically biasing the recovered flux of the GCE template.
 
\begin{figure*}[htb]
\leavevmode
\begin{center}
\includegraphics[width=.49\textwidth]{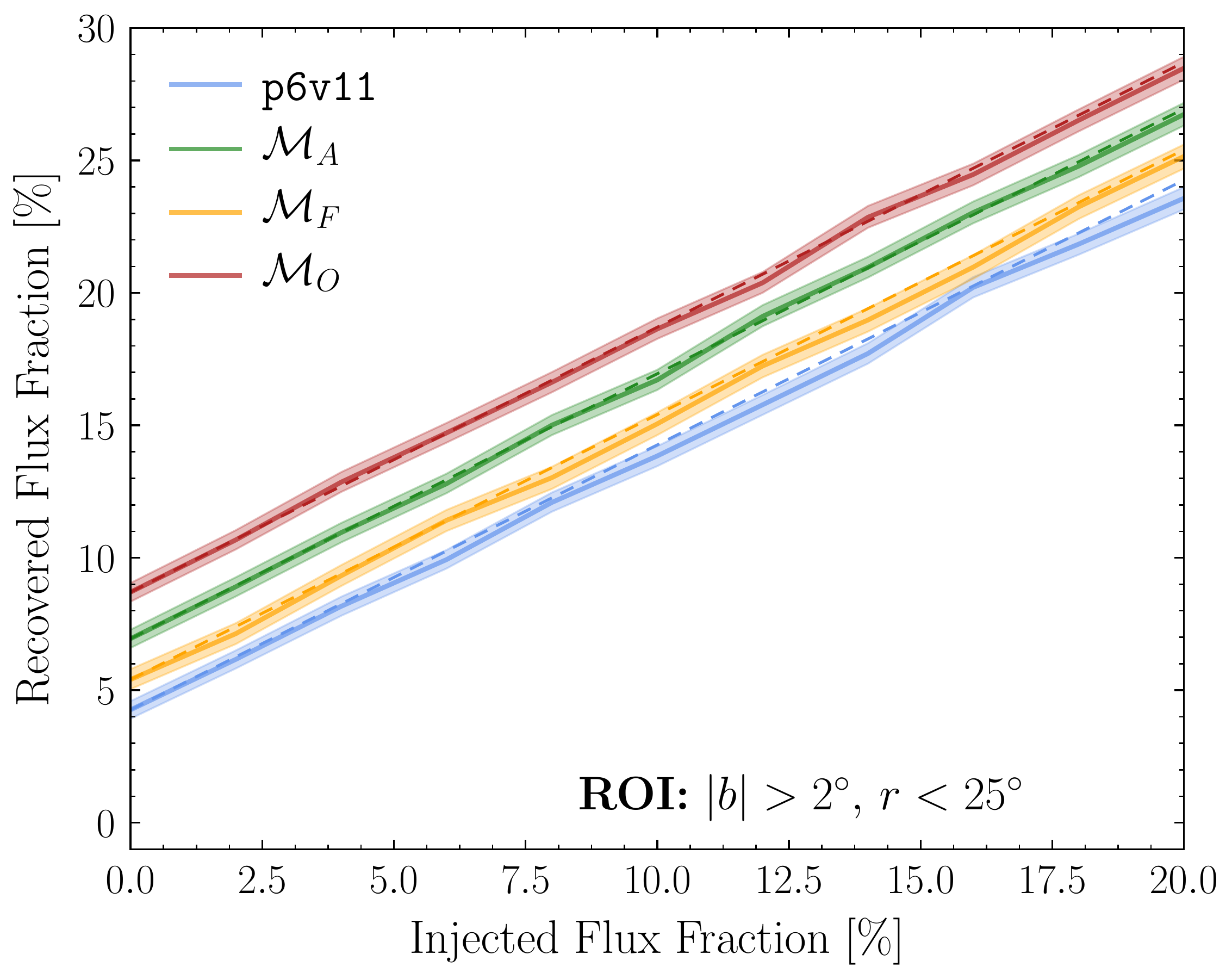}  \includegraphics[width=.49\textwidth]{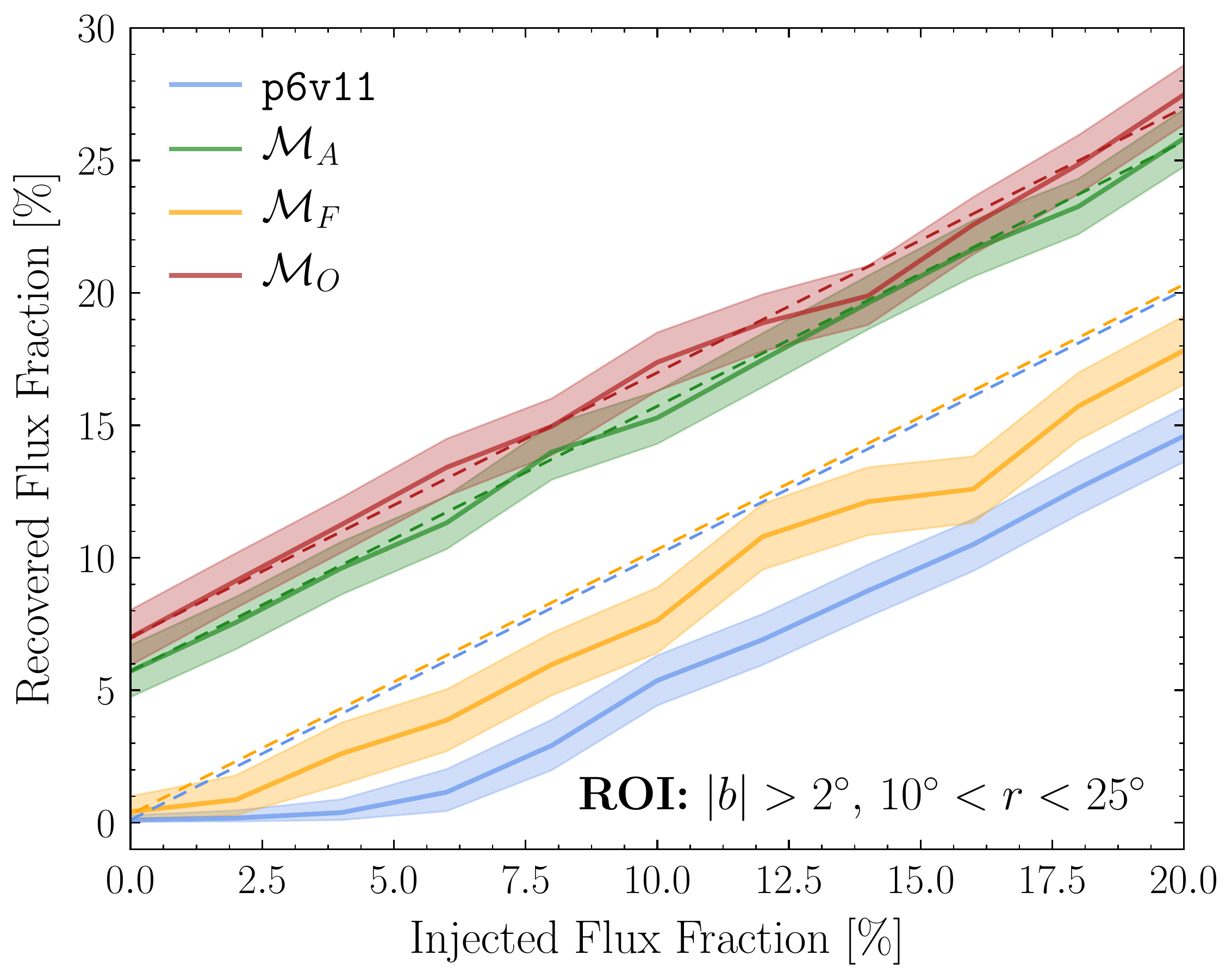} 
\end{center}
\caption{We inject an artificial DM signal on top of the real {\it Fermi} data.  The injected flux fraction is compared with the recovered flux fraction from the default Poissonian-only template analysis.  The study includes templates for diffuse emission, isotropic emission, the {\it Fermi} bubbles, 3FGL PSs, and a fiducial GCE template (modeled assuming an NFW profile).  Unlike in Figs.~\ref{fig:GCE-spec} and~\ref{fig:GCE-spec-outer}, the analyses are now run for a single energy bin that extends from 2--20~GeV.  We consider four different diffuse models: \texttt{p6v11}, as well as Models A ($\mathcal{M}_A$), F ($\mathcal{M}_F$), and O ($\mathcal{M}_O$).  These are respectively designated by blue, green, orange, and red lines/shading. We expect that the recovered flux should include both the actual GCE flux, as well as any additional injected flux; the expectations are shown by the dashed lines.  The means of the posteriors recovered by the template analysis are shown by the solid lines, with the 68\% confidence interval indicated by the shaded band. (Left)~Results for the fiducial ROI ($b > 2^\circ$ and $r<25^\circ$).  In this case, the recovered flux fractions match the expected ones.  (Right)~Results for the reduced ROI ($b > 2^\circ$ and $10^\circ < r<25^\circ$).  In this case, Model~F and \texttt{p6v11} do not produce consistent results.  This behavior is very similar to what was observed by Leane~and~Slatyer~\cite{Leane:2019uhc}, except for pure-Poissonian template fits rather than the NPTF.  We see that the artificial DM flux is not properly recovered by the analysis until a large enough flux is injected that it becomes statistically favorable for the GCE template to begin absorbing the flux again.  This behavior is  due to the fact that the \texttt{p6v11} and Model~F templates exhibit clear over-subtraction issues in this ROI, as demonstrated in Fig.~\ref{fig:GCE-spec-outer}.  Note that the flux fractions in both panels are normalized to the fiducial ROI, which has 3FGL sources masked.
}
\label{fig:DM-inject-poiss}
\end{figure*} 

The results shown in Fig.~\ref{fig:GCE-spec-outer} serve as a warning for any GCE study performed with \texttt{p6v11} (or Model~F), with 3FGL sources masked.  This diffuse model drives the GCE normalization negative in the outer regions of the Inner Galaxy, while the inner regions drive the normalization positive. As a result, the combination of best-fit model templates will necessarily over- and under-predict the data at various points when fitting over the full ROI.  When non-Poissonian templates are included, this can potentially bias the evidence in favor of PSs for the GCE, because these fluctuations can be captured to some extent by the non-Poissonian templates. 
Of course, it is important to note that not even Model O provides a description of the data at the level of Poisson noise (see App.~\ref{app:qual}). 
As such, over/under-subtraction is invariably occurring at some degree in all the diffuse models we consider.
Nevertheless, as we will demonstrate in the next subsection, when over-subtraction is demonstrably present, harmonic marginalization can alleviate it.
We therefore expect the procedure to help alleviate this systematic uncertainty more generally, even in cases where it remains undiagnosed.

One of the primary points made by Leane and Slatyer~\cite{Leane:2019uhc} is that injecting an artificial DM signal into the {\it Fermi} data and then applying the NPTF procedure produces an overly restrictive posterior for the DM template that rules out the injected  signal.  We demonstrate that this can be understood by over-subtraction by the diffuse model.  Here, we will explicitly show how such over-subtraction affects signal injection tests on data in the context of a pure Poissonian template analysis.  We will address the signal injection tests for the NPTF in the following section.   

To perform the signal injection test, we begin by summing the best-fit templates from the individual energy bins to obtain single Poissonian templates for all model components that cover the energy range from 2--20 GeV.  We also sum the {\it Fermi} data over this same energy range, so that we are only working with a single energy bin.  This is meant to facilitate comparisons with the NPTF results discussed later, which only apply to a single energy bin.  We perform a Poissonian template fit to search for evidence of the GCE, but as in Leane and~Slatyer~\cite{Leane:2019uhc}, we restrict the GCE prior to be strictly non-negative.  Moreover, we inject increasing amounts of artificial DM flux into the real {\it Fermi} data.  The results are shown in Fig.~\ref{fig:DM-inject-poiss} for the different diffuse models.

The left panel of Fig.~\ref{fig:DM-inject-poiss} shows the injected signal test for the fiducial ROI ($|b| > 2^\circ$ and $r<25^\circ$).  On the $x$-axis is the injected DM flux fraction, normalized relative to the total number of observed counts in the  ROI.  On the $y$-axis is the recovered flux fraction.  Even at zero injected flux fraction, we still recover a non-zero flux fraction because of the presence of the GCE.  As we increase the injected flux fraction, we expect to recover the original flux fraction plus whatever artificial flux is added in.  The expected flux fractions are shown by the dashed lines in Fig.~\ref{fig:DM-inject-poiss} with colors corresponding to the different diffuse models.  The solid curves are the means of the posteriors recovered on the data, with 68\% containment intervals shaded.  For the four diffuse models tested, the expected flux fractions are consistent with the results recovered on data.  Note that the recovered flux fractions vary significantly between different diffuse models because the overall normalization of the GCE is different for each model.

\begin{figure*}[htb]
\leavevmode
\begin{center}
\includegraphics[width=.49\textwidth]{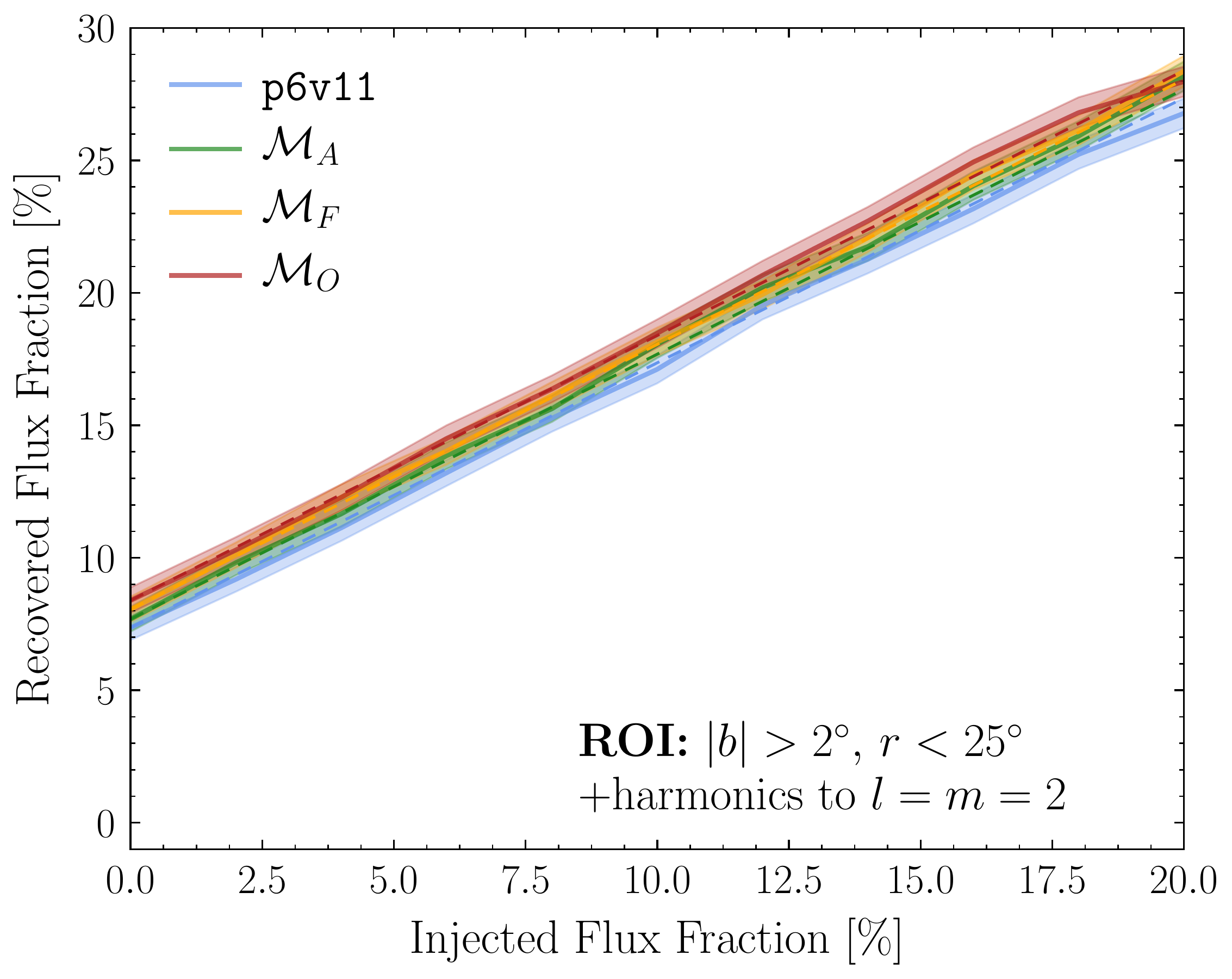}  \includegraphics[width=.49\textwidth]{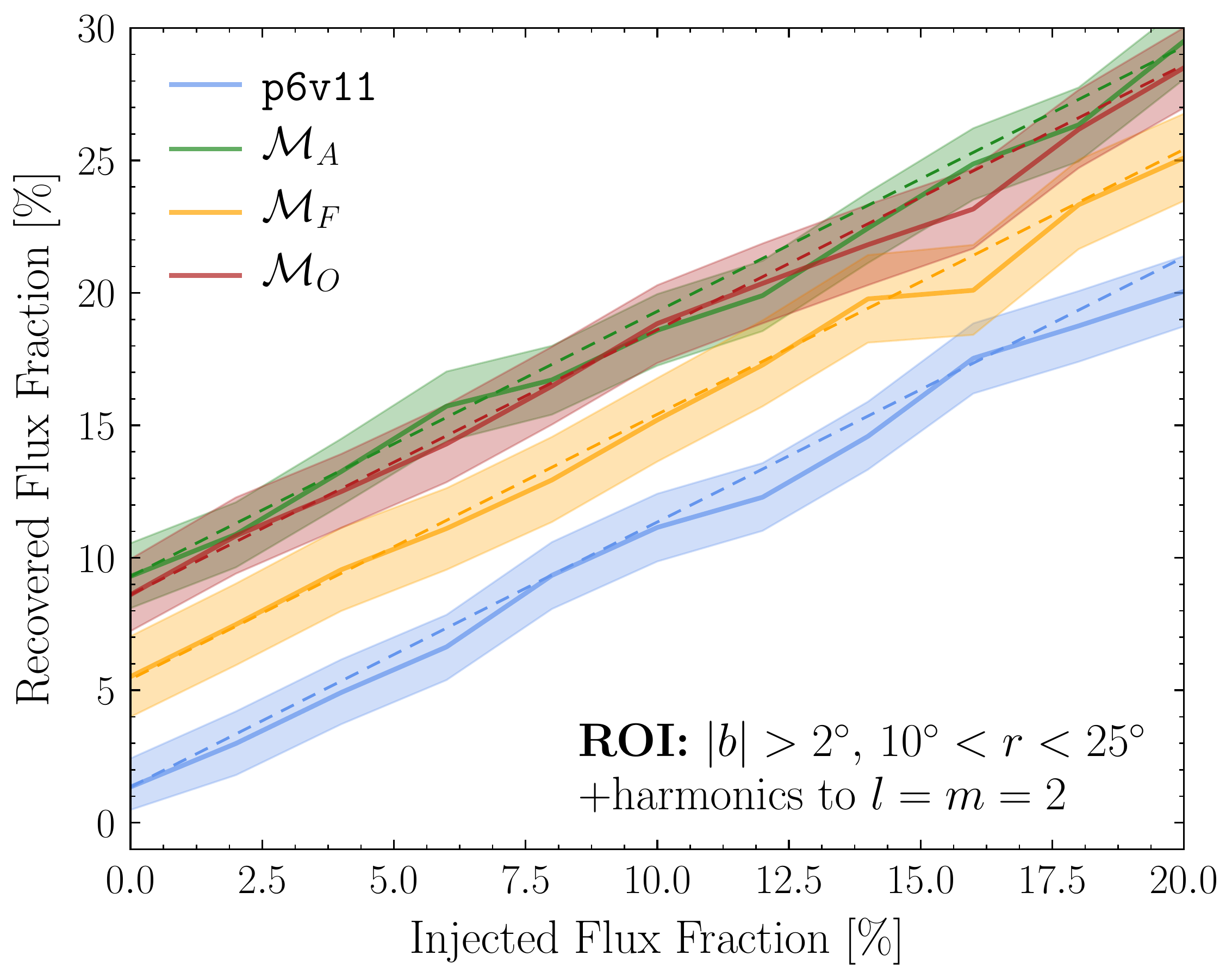} 
\end{center}
\caption{
As in Fig.~\ref{fig:DM-inject-poiss}, except that we now marginalize over harmonic templates associated with the gas-correlated components of the diffuse models out to $(\ell,m) = (2,2)$. Note that for \texttt{p6v11}, the harmonic templates are based on the entire diffuse model, which includes the IC component, while for the other models, we use the gas-correlated components as the harmonic base templates.  The recovered flux fractions are now essentially consistent with expectations in the fiducial ROI (left panel) and the reduced ROI (right panel) for all four diffuse models.  This demonstrates that harmonic marginalization can be a useful tool for mitigating the effects of diffuse mismodeling in template analyses of {\it Fermi} data.}
\label{fig:DM-inject-poiss-harm}
\end{figure*} 

In the right panel of Fig.~\ref{fig:DM-inject-poiss}, we repeat this same test in the reduced ROI where we mask the inner 10$^\circ$, bearing in mind the explicit over-subtraction we observed in Fig.~\ref{fig:GCE-spec-outer}.
The Model~A and~O results behave as expected in this ROI, with the recovered flux fractions following the expectation.
However, the \texttt{p6v11} and Model~F results are not consistent with the expectations.  At zero injected flux, the recovered fraction is zero, but as small amounts of GCE flux are added to the data, the recovered flux remains zero. The failure of the \texttt{p6v11} and Model~F cases to pass this test is related to the fact that the recovered flux fractions really want to be negative at the $\sim$5\% level because of over-subtraction, and would float to these values if the priors allowed it (see Fig.~\ref{fig:GCE-spec-outer}).
Due to this, the injected flux fraction needs to be above $\sim$5\% before a positive flux fraction is again preferred.

The results of Fig.~\ref{fig:DM-inject-poiss} (right panel) are the equivalent of the  results presented in Leane and Slatyer~\cite{Leane:2019uhc}, except for purely Poissonian template fits.  We clearly see that artificial DM signals injected in the data are not properly recovered by the template analysis, until some large enough flux is added in.  Based on the behavior observed in the recovered spectra (Fig.~\ref{fig:GCE-spec-outer}), we believe that this is due to the fact that the injected DM signal is absorbed by the (poorly modeled) diffuse foreground template, up until it becomes statistically favorable for the GCE template to begin absorbing the injected flux.

\subsection{Harmonic Marginalization and Over-subtraction}

We now investigate how the harmonic marginalization procedure described in Sec.~\ref{sec:harm} can help mitigate the over-subtraction issue illustrated in the right panel of Fig.~\ref{fig:DM-inject-poiss}.  Specifically, we add in harmonic nuisance templates that are derived from the diffuse template up to and including the $(\ell,m) = (2,2)$ mode.  That is, our original diffuse template is replaced by nine templates that are multiplied by all the harmonics through $\ell = 2$, according to the procedure described in Sec.~\ref{sec:harm}.  After injecting a synthetic DM signal into the data, we recover the flux fractions after marginalizing over the nuisance parameters.  Figure~\ref{fig:DM-inject-poiss-harm} summarizes the results of the harmonic marginalization analyses.   The left panel presents the results for the fiducial ROI.  We see that the results for the different diffuse models are now more consistent with each other, as compared to the left panel of Fig.~\ref{fig:DM-inject-poiss}.  Indeed, Models~A, F, and O give nearly identical results, and \texttt{p6v11} has a recovered flux fraction that is only $\sim$$10\%$ lower than the others.

The right panel of Fig.~\ref{fig:DM-inject-poiss-harm} focuses on the case of the reduced ROI, with the inner $10^\circ$ masked, where the over-subtraction issues for \texttt{p6v11} and Model~F are particularly apparent.  With the harmonic analysis, this issue is now essentially resolved for both, with the expected and observed flux fractions now consistent with each other. This demonstrates that the harmonic marginalization procedure is able to partially mitigate over-subtraction from diffuse mismodeling. We emphasize that of course in the absence of a perfect diffuse model, mismodeling remains. Nevertheless, these results demonstrate harmonic marginalization has enhanced the robustness of fundamental properties extracted for GCE to such systematics.

Note that for Models~A, F, and~O, we assign independent nuisance parameters to the gas-correlated and IC templates, while for \texttt{p6v11} these contributions are summed into one template with only a single nuisance parameter.  In Fig.~\ref{fig:DM-inject-poiss-harm}, we performed the harmonic marginalization procedure for the entire \texttt{p6v11} template, while for the other diffuse models, we only performed harmonic marginalization on the gas-correlated templates.  Adding additional harmonic modes for the IC templates, or indeed other templates such as the isotropic component, leads to a slightly better correspondence between diffuse models.
Nevertheless, as the greatest improvement is associated with adding harmonics to the gas-correlated maps, we restrict our attention to these for simplicity.

The tests in this section give us confidence that the harmonic marginalization procedure is able to partially mitigate the effects of diffuse mismodeling on the GCE.  Armed with this new method, we now turn to the implications for the evidence of unresolved PSs in the Inner Galaxy using the NPTF. 

\section{spherical-harmonic Marginalization and the NPTF}
\label{sec:harmNPTF}

In this section, we explore how the harmonic marginalization procedure may be combined with the NPTF to reduce the systematic uncertainties that arise when inferring the presence and properties of an unresolved PS population in the Inner Galaxy.  We begin with a toy MC example before turning to the actual {\it Fermi} dataset.  We then conclude by performing the same signal injection tests done by Leane~and~Slatyer~\cite{Leane:2019uhc}, but with an improved treatment of the diffuse foreground modeling.  We demonstrate that the primary issue pointed out in that work---namely, that an artificial DM signal injected into the data is not properly recovered by the NPTF---is due to the choice of diffuse foreground models used.  As we show, when the treatment of diffuse models is improved, artificial DM signals are properly recovered, and the evidence for PSs remains robust, under the variations that we have tested.

\subsection{A Toy Example in Simulated Data}
\label{sec:harm-toy}

We begin by building a set of simulated data maps that can be run through the NPTF analysis pipeline.  This exercise will allow us to test whether the harmonic marginalization procedure can effectively mitigate the effects of diffuse mismodeling.  
The simulated data is created using Model~O, since this is the model---amongst the four considered here---that provides the best fit to the data.  When analyzing the simulated data maps with the NPTF, we have the freedom to select whatever diffuse template we like.  If we use a template based on Model~O, then by default the analysis assumes no uncertainty in the foreground modeling.   This is the best-case scenario, though one we know is unrealistic for actual {\it Fermi} analyses.  To mock-up the effects of diffuse mismodeling, which we know play an important role on the actual data, we can choose instead to use a template that is not based on Model~O.  In this subsection, we will consider both cases, using both the \texttt{p6v11} and the Model~O templates in analyses.  
 
The best-fit template normalizations used to construct the simulated data are obtained from an NPTF fit in the ROI with $|b| < 2^\circ$, $r < 15^\circ$, and 3FGL PSs masked and in the single energy bin (2--20~GeV).\footnote{We find consistent GCE PS results if instead we use the template normalizations and source-count distribution model parameters from a fit in our fiducial ROI.  The reduced ROI is used here, however, in order to minimize the possible impact of bright disk-correlated PSs extending beyond the 3FGL mask and biasing the disk-correlated source-count distribution.} We show the source-count distributions recovered for various choices of the diffuse models in App.~\ref{app:extended}. The Poissonian emission includes the following components: the gas-correlated and IC emission from Model~O, isotropic emission, and the {\it Fermi} bubbles. The non-Poissonian emission includes: disk-correlated PSs and GCE-correlated PSs.\footnote{We do not include an isotropic PS template because we find that it does not significantly improve the fit to the real data in the fiducial ROI.}  Note that we do not include a GCE Poissonian template in the data fit to get the template normalizations, although some of the MC generated includes a Poissonian GCE component. 
Explicitly, we generate two different classes of simulated data: \emph{(i)}~we include GCE-correlated PSs, with best-fit model parameters as found in the NPTF analysis of the real data, and \emph{(ii)}~we include GCE-correlated Poissonian emission, with total flux matching that recovered for  GCE-correlated PSs from the analysis of the real data.
In both cases we also simulate disk-correlated PSs. The two classes of MC are designed to model the situations where the GCE arises purely from PS or smooth emission.

\begin{figure}[t]
\leavevmode
\begin{center}
\includegraphics[width=0.5\textwidth]{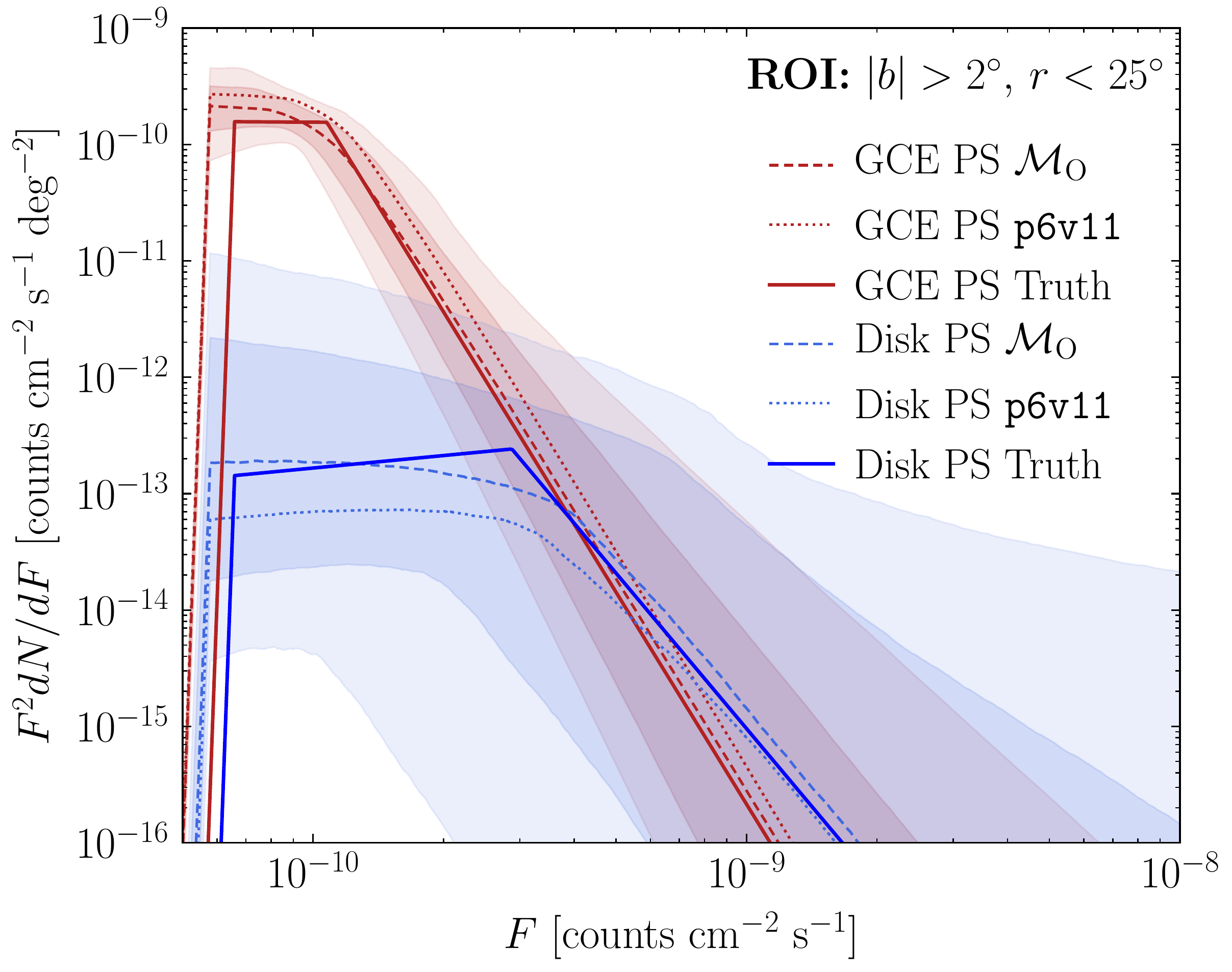} 
\end{center}
\caption{Source-count distributions for the unresolved GCE- and disk-correlated PS populations, in red and blue, respectively.  The solid lines show the best-fit distributions recovered from an NPTF analysis on {\it Fermi} data in the fiducial ROI and energy range, using the Model~O diffuse templates.  These are taken as the `truth' distributions when generating the simulated data maps.   
We create the simulated data maps assuming that the diffuse emission traces Model~O.  We then run the NPTF analysis pipeline on the simulated data, considering two different diffuse model scenarios.  In the first case, labeled as $\mathcal{M}_O$, we include the gas-correlated and IC emission templates from Model~O in the NPTF.  The best-fit source counts that are recovered are shown by the dashed lines, with the shaded bands denoting the 68\% and 95\% containment regions.  For the same simulated dataset, we repeat the NPTF analysis using the \texttt{p6v11} template.  This is intended to mock-up the effects of diffuse mismodeling.  The best-fit results are shown by the dashed lines (containment regions not shown).  The recovered source-count distributions for both scenarios are consistent with the truth distributions, though this need not necessarily be the case for other manifestations of diffuse mismodeling.  The source-count distributions are normalized to the fiducial ROI and energy range.
}
\label{fig:NPTF-MC-dNdF}
\end{figure}
 
The source-count distributions for the GCE (red) and disk-correlated (blue) PSs are shown by the solid lines in Fig.~\ref{fig:NPTF-MC-dNdF} (labeled as `Truth').  Note that these source-count distributions are normalized for the fiducial ROI and energy range.  The resolution threshold for 3FGL sources is approximately $F \sim 3 \times 10^{-10}$~counts~cm$^{-2}$~s$^{-1}$ in this energy range and ROI, though the actual detection threshold depends on spatial location and spectral shape.  When the 3FGL sources are not masked in the  analysis, then we find that the disk-correlated source-count distribution has support at high flux, consistent with the observed 3FGL flux distribution.
As we see, the fit prefers more GCE-correlated sources than disk-correlated sources below the 3FGL threshold.  The hard lower cutoff in the truth distributions at $F \sim 6 \times 10^{-11}$~counts~cm$^{-2}$~s$^{-1}$ is due to the fact that we force the source-count distribution to zero below a certain point in order to reduce the effect of a potential degeneracy between ultrafaint sources and smooth emission (from \emph{e.g.}, DM) with the same spatial distribution.  In practice, we determine the lower-flux cutoff through the following procedure, which roughly approximates the typical 1-$\sigma$ detection threshold for PS detection through traditional algorithms.  We first determine the average number of counts within a PSF radius with the ROI.  The typical 1-$\sigma$ PS detection threshold, in terms of counts, is then given by the square root of this number.  We convert this number to flux using the average exposure within the ROI and then set this flux value as the lower cutoff of the source-count distribution.

\begin{figure*}[htb]
\leavevmode
\begin{center}
\includegraphics[width=1.0\textwidth]{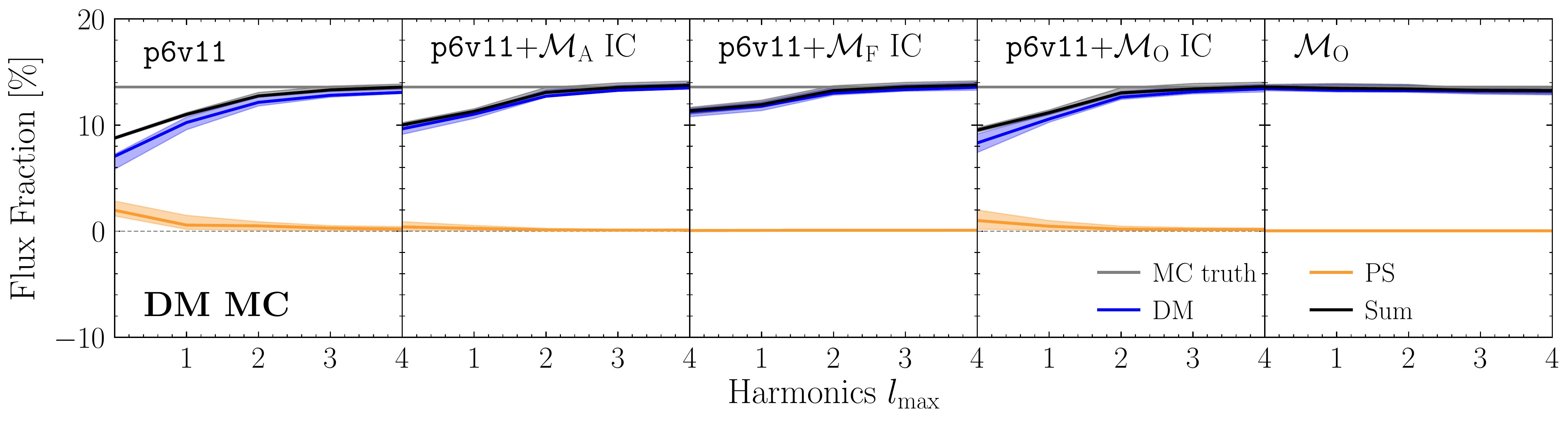} 
\includegraphics[width=1.0\textwidth]{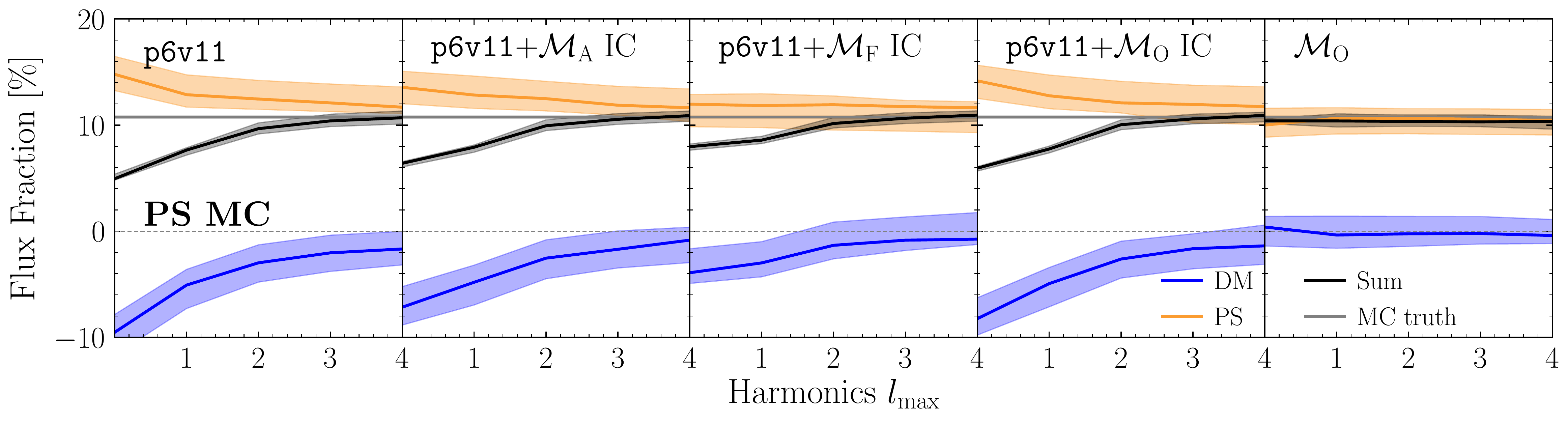} 
\end{center}
\caption{Flux fractions recovered from the simulated data when the GCE arises  entirely from DM annihilation (top row) or from a population of unresolved PSs (bottom row).  All simulated datasets are generated assuming diffuse emission that traces Model~O.  Each column corresponds to a different diffuse template(s) used in the NPTF analysis: \texttt{p6v11}, \texttt{p6v11} + IC components of Model~A, F, and O, and Model~O.  Cases where the NPTF assumes a diffuse model that is different than Model~O are intended to mock-up instances of foreground mismodeling, which are expected on actual data.  When \texttt{p6v11} is supplemented with an additional IC map, the normalization of the latter is allowed to scan negative.  The results in each panel are provided as a function of the maximum spherical-harmonic number $\ell_{\rm max}$ that is marginalized (see text for details).  $\ell_\text{max} =0$ corresponds to the standard case where no harmonic marginalization is performed.  The recovered flux fractions for the GCE-Poissonian and GCE-PS templates are provided in blue and orange, respectively, and their sum is shown in black.  The solid lines denote the $50^\text{th}$ percentile (and the bands indicate the 16 and $84^\text{th}$ percentiles) over eleven Monte Carlo iterations of the simulated map.  The thick solid gray line indicates the true flux fraction, which is taken to be 13.6\% for the DM MC and 10.8\% for the PS MC. When the GCE consists entirely of PSs and the \texttt{p6v11} template is used in the NPTF analysis, there is significant over-subtraction when $\ell_{\rm max} = 0$, whereby the DM template recovers a negative flux fraction.  As we marginalize over increasing $\ell_{\rm max}$, the over-subtraction, which is due to mismodeling the diffuse emission, is mitigated.  When the GCE consists entirely of DM and the \texttt{p6v11} template is used, then the recovered DM fraction is still low at $\ell_\text{max} =0$ and there is some flux that is absorbed by the PS template, likely from residual diffuse emission.  These results are broadly consistent across the examples presented here.  When the NPTF uses the Model~O templates, then the recovered flux fractions are more consistent with truth even at low $\ell_\text{max}$ and the harmonic marginalization has less of an effect.  This is to be expected because there is no diffuse mismodeling.
}
\label{fig:NPTF-harm-MC}
\end{figure*}

Figure~\ref{fig:NPTF-harm-MC} compactly summarizes the results of the tests that we performed on simulated data.  The top row corresponds to the case where the GCE arises from only DM, while the bottom row corresponds to the case where it consists entirely of GCE-correlated PSs (that follow the truth source-count distribution shown in Fig.~\ref{fig:NPTF-MC-dNdF}).  Each column shows the results obtained by varying the diffuse template used in the NPTF analysis; recall that the simulated data is itself generated with Model~O in all cases, therefore the last column corresponds to the case of no diffuse mismodeling.  Each panel shows the recovered flux fractions for the GCE-Poissonian (blue) and GCE-PS (orange) templates, as a function of the maximum spherical-harmonic number, $\ell_\text{max}$.   The bands are computed by taking the 16\% and 84\% percentiles of the best-fit recovered flux fractions over an ensemble of eleven independent MC realizations; the solid curves show the 50\% percentile results.  The true simulated value for the flux fraction is indicated by the thick gray line in each panel. Slightly different values are simulated for the two cases: 13.6\% for the DM MC and 10.8\% for the PS MC. Note that, for a given $\ell_{\rm max}$, we include all $(\ell, m)$ harmonic templates with lower $\ell \leq \ell_{\rm max}$.  For example, $\ell_{\rm max} = 0$ implies that we only include the $(0,0)$ harmonic template, which is equivalent to not performing the harmonic marginalization procedure at all.  The value $\ell_{\rm max} = 2$ means that we include the harmonic templates $(0,0)$, $(1,-1)$, $(1,0)$, $(1,1)$, $(2,-2)$, $(2,-1)$, $(2,0)$, $(2,1)$, and $(2,2)$. 

We begin by considering the instance where $\ell_\text{max} = 0$, which corresponds to the case of no harmonic marginalization. When using the Model~O template, we accurately recover the flux fractions of GCE-correlated PSs and DM for both simulated data maps, up to small biases pointed out in Ref.~\cite{Chang:2019ars}.  This is to be expected because the diffuse template perfectly models the diffuse emission in the simulated map.  However, the results are not as clean when we instead use the \texttt{p6v11} template.  In this case, we do not recover the correct flux fractions for the GCE-correlated emission.  In particular, when the simulated data has a DM GCE,  the analysis recovers significantly less DM than what is actually in the simulated data.  It also finds a non-zero flux fraction for GCE PSs, even though no PSs are actually present in the simulated data.  This is likely due to residuals from the diffuse mismodeling mimicking PSs.  When the simulated data instead includes a population of GCE-correlated PSs, we find that using the \texttt{p6v11} diffuse template recovers too much PS flux.  This is at the expense of the Poissonian DM template normalization being driven to unphysical negative values, a clear sign of over-subtraction.  Clearly, mismodeling the diffuse emission in the presence of unresolved PSs leads to over-subtraction of the Poissonian DM template.

 \begin{figure*}[htb]
\leavevmode
\begin{center}
\includegraphics[width=0.49\textwidth]{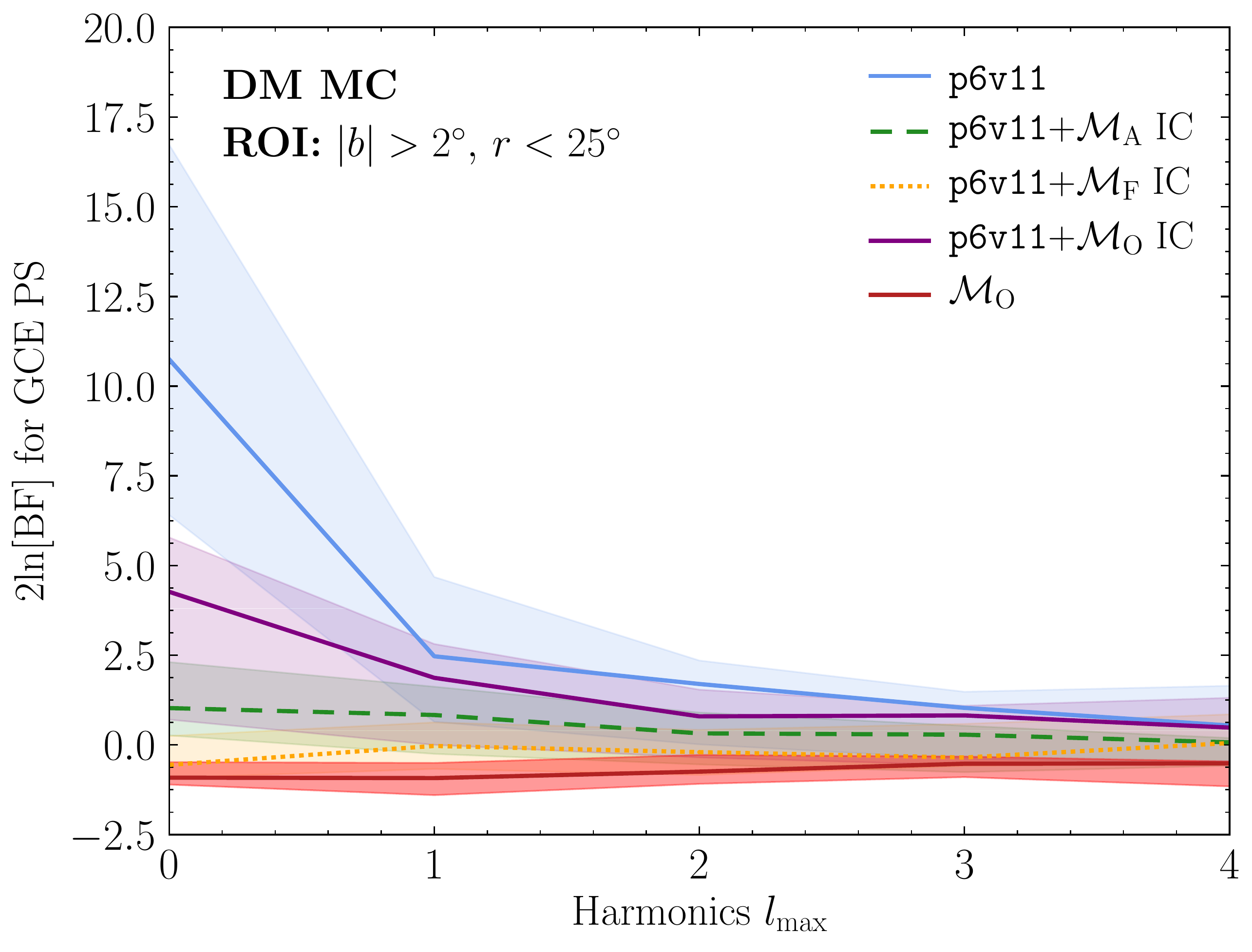} 
\includegraphics[width=0.49\textwidth]{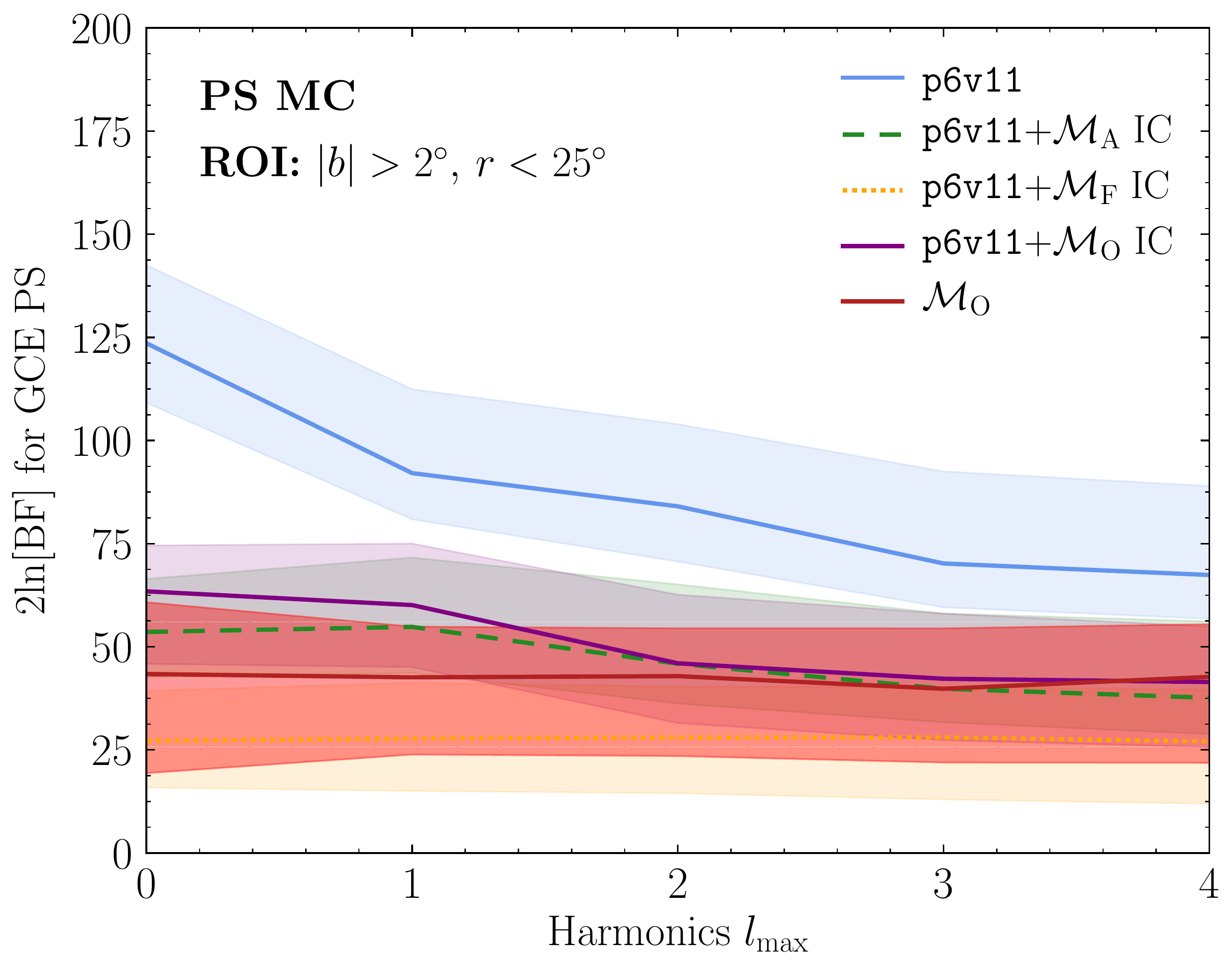} 
\end{center}
\caption{
The Bayes factors in preference for models with GCE PSs for the simulated data tests presented in Fig.~\ref{fig:NPTF-harm-MC}, except with the additional requirement that the DM flux fraction must be positive ({\it i.e.}, physical).  (Left) The results for when the simulated data has a DM GCE.  In this case, we expect a negligible Bayes factor in preference for the model with  PSs.  Indeed, when we use the true diffuse model for the template (Model~O), we find a Bayes factor less than unity, indicating no evidence for PSs.  For different diffuse templates, the evidence in favor of GCE PSs is either trivial to begin with or decreases with increasing harmonic number $\ell_{\rm max}$, as would be expected.  (Right) When the GCE consists of PSs, we robustly find evidence in favor of PSs.  However, we do caution that in this case, the \texttt{p6v11} analysis finds overly strong evidence in favor of PSs even after spherical-harmonic marginalization.  On the other hand, supplementing \texttt{p6v11} with an additional IC template (whose normalization can scan negative) brings the evidence to similar values as found with the true diffuse model (Model~O).
}
\label{fig:MC-bayes}
\end{figure*}

We may partially mitigate the over-subtraction simply by supplementing the \texttt{p6v11} diffuse model with an additional IC template.  One central difference between how we implement {\it e.g.}, the Model~O and \texttt{p6v11} diffuse models is that the former has two different components, one for IC and the other for gas-correlated emission, while the latter has both of these components already summed together.  Thus, by supplementing the \texttt{p6v11} model with an IC model---and allowing the normalization of this model to scan negative---we are able to correct for any potential differences in relative normalization between the gas-correlated and IC components.  We supplement \texttt{p6v11} with three different IC templates to assess the importance of mismodeling the IC emission.  In particular, we use the IC templates from Models~A, F, and~O.  We find that supplementing \texttt{p6v11} with these IC templates partially mitigates the issues described above, but that over-subtraction persists when $\ell_\text{max} = 0$.

Next, we explore how the spherical-harmonic marginalization procedure can mitigate the bias induced from using the wrong diffuse model.  We begin by performing the spherical-harmonic marginalization procedure on either the gas-correlated template for Model~O or the \texttt{p6v11} template, depending on which is used in the analysis.  
For the Model~O case, the results are not affected by the harmonic marginalization procedure and we continue to recover the correct template normalizations for PSs and DM.  This makes sense because the diffuse template is an accurate representation of the true diffuse emission in the simulated data. 

The harmonic marginalization procedure plays a significant role, on the other hand, when the \texttt{p6v11} template is used.  In the case where the GCE arises from DM, for example, we see that marginalizing the harmonic templates leads to the DM posterior approaching the true  value. This result is robust to the choice of IC model that supplements the \texttt{p6v11} template and is even true when no extra IC template is present.  Similarly, in all cases, we see that when the GCE arises from PSs, performing the spherical-harmonic marginalization procedure mitigates the over-subtraction and the DM template normalization goes from being negative to closer to zero.

It is instructive to also compute the Bayes factor between the model with spherical PSs and DM versus that without spherical PSs, for both simulated datasets.  For this computation, we restrict the prior on the DM normalization to only cover positive ({\it i.e.}, physical) values so that we are comparing two physical models.  The Bayes factor comparisons are shown in Fig.~\ref{fig:MC-bayes}.
The left panel shows the Bayes factors when the simulated data has a DM GCE, while the right panel shows the results when the GCE arises from PSs.  In each case, we show how the Bayes factors change with harmonic number in the spherical marginalization procedure, for different choices of the diffuse templates.  The bands arise from performing eleven MC analyses and taking the 16\% and 84\% percentiles of the Bayes factor distributions.

When the GCE arises from DM and the data is analyzed with the correct diffuse template (Model~O), the Bayes factor in preference for GCE-distributed PSs is less than unity, indicating that the GCE PSs are not favored.  On the other hand, when the GCE does arise from PSs and the Model~O template is used, the evidence in favor of PSs is encapsulated by a Bayes factor around $2 \ln {\rm BF} \sim 40 \pm 15$. This demonstrates that, at least when the true diffuse model is used in the template analysis, the Bayes factor is an effective diagnostic for providing evidence for or against GCE-correlated PSs.  Moreover, the Bayes factors in each case are relatively insensitive to the maximum harmonic number in the marginalization procedure. 

\begin{figure*}[htb]
\leavevmode
\begin{center}
\includegraphics[width=1.0\textwidth]{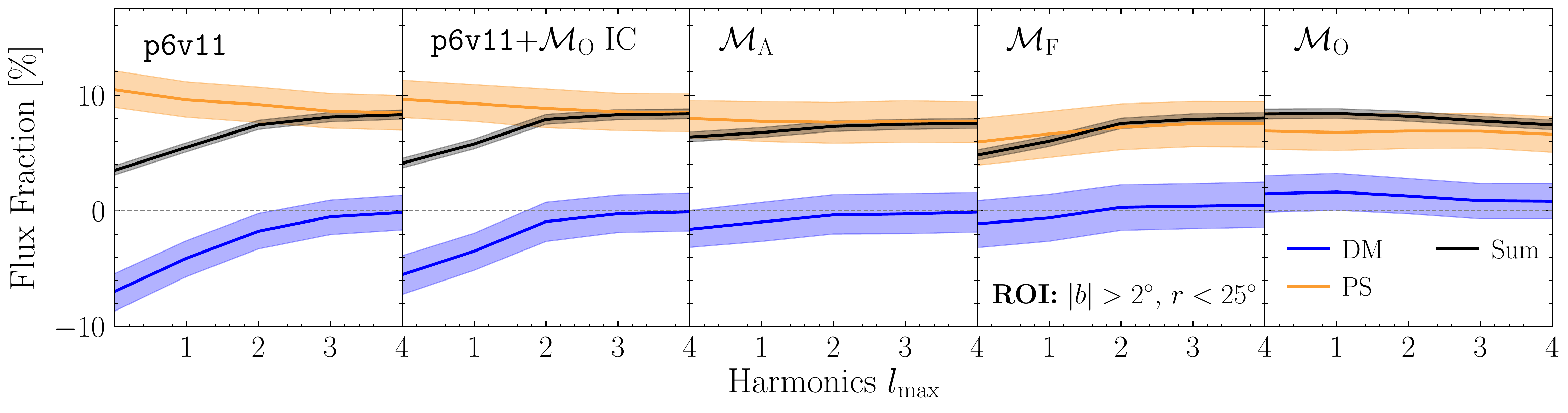} 
\end{center}
\caption{
As in Fig.~\ref{fig:NPTF-harm-MC}, except for the actual {\it Fermi} data.  We show the recovered flux fractions for the DM, GCE-correlated PS, and DM + PS templates when the different diffuse models (as indicated) are used in the NPTF analysis.  For \texttt{p6v11}, there is clear over-subtraction at $\ell_\text{max} = 0$, as evidenced by the negative flux fractions for the DM template.  The spherical-harmonic marginalization procedure mitigates the over-subtraction, when it is present.  In all cases, by large $\ell_\text{max}$, we recover a DM posterior that is consistent with zero flux and a robust, non-zero flux for the GCE-correlated PSs.}
\label{fig:NPTF-harm-data}
\end{figure*}

When the incorrect diffuse model is used in the template  analysis, the harmonic marginalization procedure is able to effectively reduce the artificial Bayes factor found in preference for GCE PSs when the simulated data is constructed with a DM GCE.  This is seen in the left panel of Fig.~\ref{fig:MC-bayes}.  For example, when only the \texttt{p6v11} diffuse model is used in the analysis, the Bayes factor in preference for GCE-correlated PSs is around $10^2$ without performing the harmonic marginalization, even though there are no GCE-correlated PSs in this case.  However, after performing the harmonic marginalization procedure, the Bayes factor decreases towards a negligible value.  We also see that including a separate IC template in conjunction with the \texttt{p6v11} model reduces the  evidence for PSs.  When the GCE is constructed from spherical PSs, the Bayes factors can be artificially enhanced if the wrong diffuse template is used in the analysis.  This is particularly pronounced for the \texttt{p6v11} template.  In this case, the harmonic marginalization procedure does reduce the Bayes factors slightly, though it still does not reach the levels found with Model~O.  However, including the IC template in conjunction with \texttt{p6v11}, regardless of which IC template, does produce Bayes factor results in good agreement with those found with Model O.

Lastly, it is interesting to compare the recovered source-count distributions from the analyses of the simulated data with PSs to the true source-count distributions used in creating the simulated data.  These comparisons are shown in Fig.~\ref{fig:NPTF-MC-dNdF}.  We compare example source-count distributions from analyses that use the Model~O templates and those that use the \texttt{p6v11} template.  In both cases, we accurately recover the shapes of the source-count distribution.  It is interesting to note, and somewhat surprising, that even when using the \texttt{p6v11} template, we are able to accurately recover the source-count distribution (within uncertainties), which suggests that at least for this test, the shape of the source-count distribution is less subject to bias from diffuse mismodeling than {\it e.g.}, the flux fractions and the evidence. 

\begin{figure}[!htbp]
\leavevmode
\begin{center}
\includegraphics[width=0.49\textwidth]{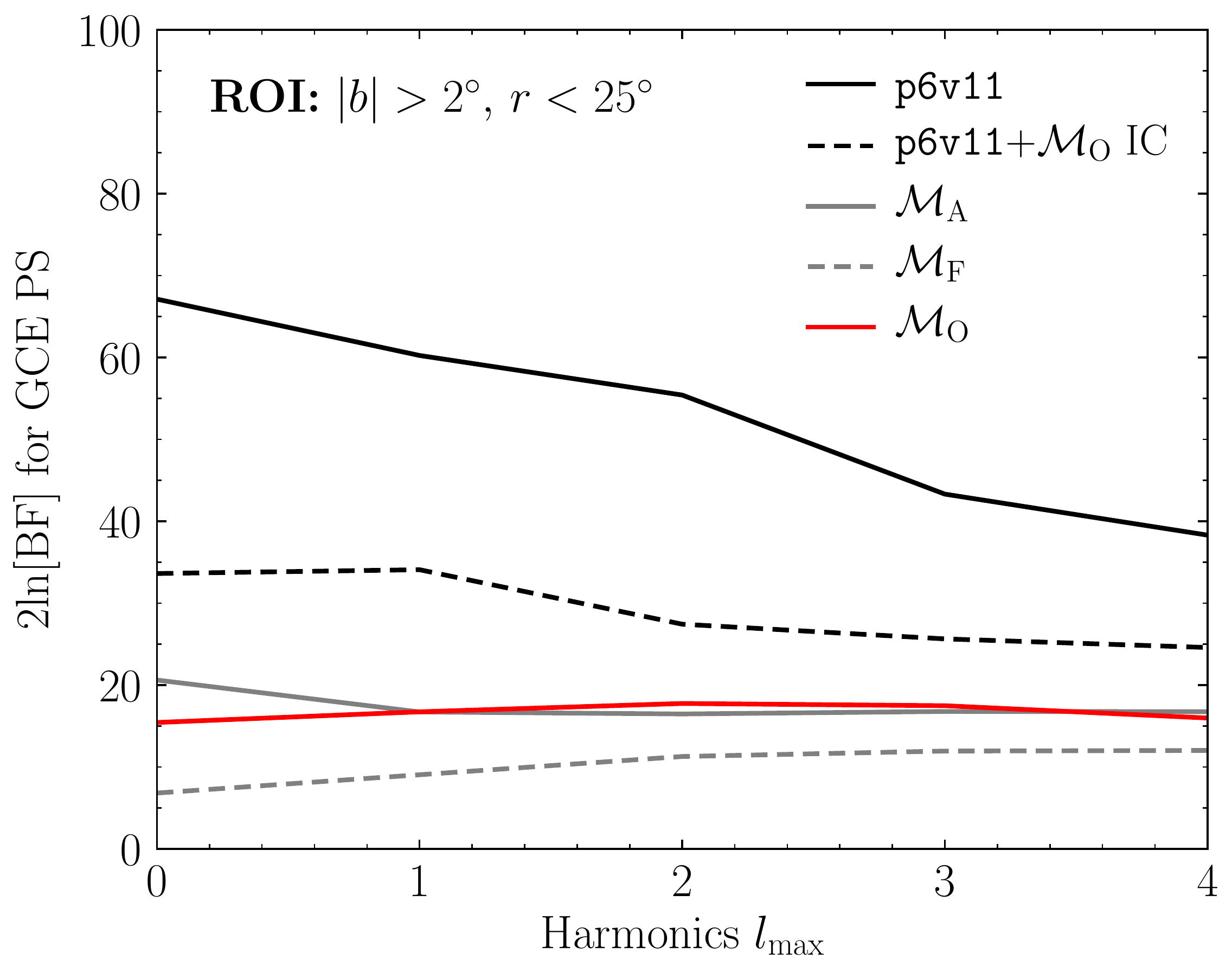} 
\end{center}
\caption{The Bayes factor in preference for the model with PSs over that without for an analysis of the real {\it Fermi} data in the fiducial ROI and energy range.  The examples provided parallel those in Fig.~\ref{fig:NPTF-harm-data}.  We marginalize over increasing harmonic numbers $\ell_{\rm max}$.  In all cases we find evidence for spherical PSs over DM for the GCE. 
}
\label{fig:NPTF-bayes-data}
\end{figure}


\subsection{Application to {\it Fermi} Data}

\begin{figure*}[htb]
\leavevmode
\begin{center}
\includegraphics[width=0.49\textwidth]{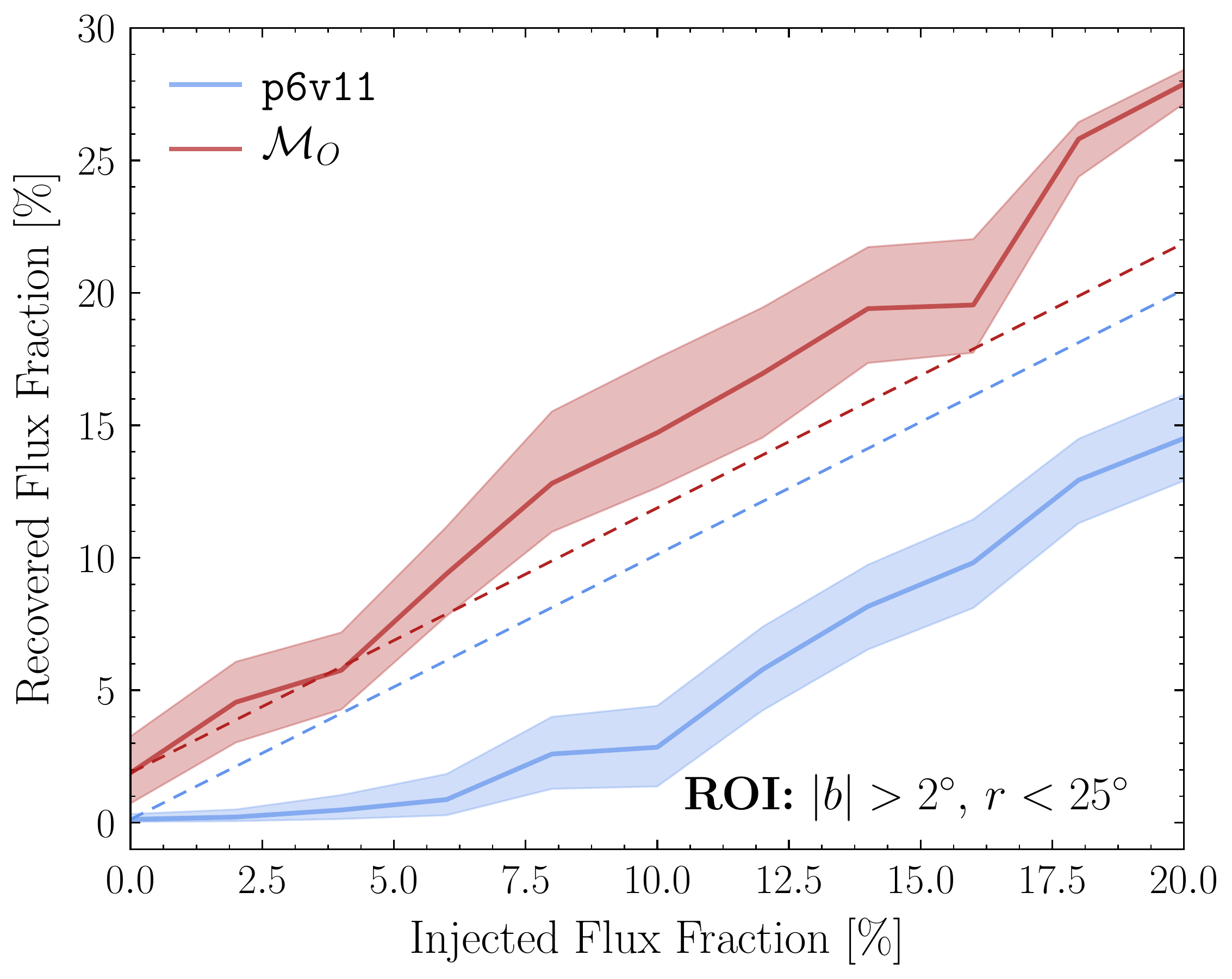} 
\includegraphics[width=0.49\textwidth]{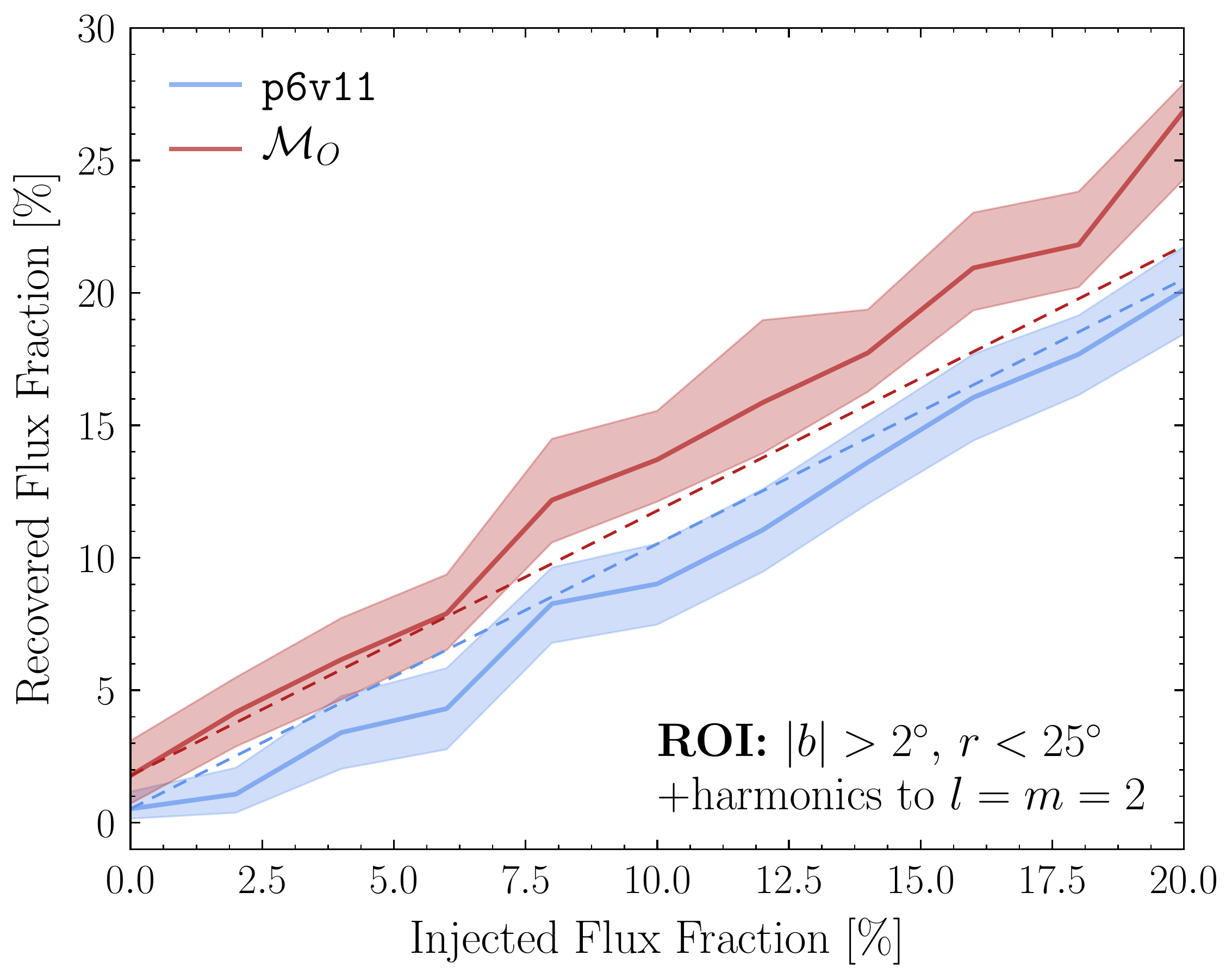} 
\end{center}
\caption{Results of signal injection tests whereby a synthetic DM signal is injected into the real {\it Fermi} data. (Left) The hybrid data is analyzed with the NPTF including both a Poissonian GCE model and a non-Poissonian GCE-correlated PS component.  The Poissonian GCE model is not allowed to float negative.  The dashed curves show the expected recovered flux fractions for each diffuse model, while the bands show the 68\% containment intervals from the posteriors for the analyses of the hybrid data using both the \texttt{p6v11} and Model~O.  \texttt{p6v11} shows clear issues with over-subtraction; the recovered flux fraction is consistently lower than the expectation.  For Model~O, where mismodeling effects are less pronounced, the recovered flux fractions are more consistent with expectation.  (Right)~As in the left panel, but now the diffuse models are improved by performing the spherical-harmonic marginalization procedure through $\ell = m = 2$.  This has a dramatic effect in bringing the \texttt{p6v11} results (and, to a lesser extent, Model~O results) in line with the expectations.  This clearly demonstrates that using improved diffuse emission models, such as Model~O, or implementing measures such as spherical-harmonic marginalization resolves the anomalous signal-injection test results reported by Leane~and~Slatyer~\cite{Leane:2019uhc}. }
\label{fig:NPTF-harm-data-sig-inj}
\end{figure*}

Next, we repeat the spherical-harmonic marginalization procedure described above on the actual {\it Fermi} data, using the fiducial ROI and energy range.  The results of this test are summarized in Fig.~\ref{fig:NPTF-harm-data}.  This figure shows the recovered flux fraction as a function of the harmonic number $\ell_{\rm max}$ for different diffuse templates, just as in Fig.~\ref{fig:NPTF-harm-MC}.  Importantly, the solid lines in Fig.~\ref{fig:NPTF-harm-MC} now correspond to the centers of the posteriors for the specific model components, with bands indicating the 68\% containment regions as computed from the posterior. 

When $\ell_\text{max} = 0$, there is over-subtraction in the Poissonian DM template when using \texttt{p6v11}.  This is similar for the \texttt{p6v11}+$\mathcal{M}_\text{O}$~IC case. In these cases, however, the spherical-harmonic marginalization procedure mitigates this over-subtraction as we marginalize over increasingly large $\ell_{\rm max}$.  Interestingly, the Models~A,~F, and~O results seem to be relatively insensitive to the harmonic marginalization procedure for the ROI considered, which we take as evidence that these models are a comparatively good description of the underlying diffuse emission.  
In each case, after accounting for the harmonic marginalization, we find that the DM posterior is consistent with zero smooth DM flux at 68\% confidence and that the GCE PS posteriors give consistent and non-zero results.  This provides evidence that the GCE is better explained by PS-like emission than by smooth emission even when issues of diffuse mismodeling are mitigated. 

To assess the evidence for GCE-correlated PSs over Poissonian DM at a more quantitative level, it is instructive to compute the Bayes factor between the model with GCE PSs and DM versus that without PSs, following the procedure that we outlined on simulated data. 
The resulting Bayes factors are shown in Fig.~\ref{fig:NPTF-bayes-data}.  The Model~O Bayes factor is relatively insensitive to the harmonic marginalization procedure and is approximately $\sim$$10^{3.4}$.  
The Bayes factors for the individual \texttt{p6v11} examples converge towards the Model~O results at large $\ell_\text{max}$, with the largest discrepancy remaining for the case when \texttt{p6v11} is used on its own.

It is worth commenting that with a similar dataset the Bayes factor in favor of PSs over smooth emission for the GCE was found to be $\sim$$10^4$ in Ref.~\cite{Lee:2015fea} when using the \texttt{p6v11} diffuse model and masking 3FGL sources, which is substantially smaller than the Bayes factor found here when using \texttt{p6v11} for a similar analysis.  We believe that much of this discrepancy comes from the difference in 3FGL PS mask used between our present work and Ref.~\cite{Lee:2015fea}.  The 3FGL mask in Ref.~\cite{Lee:2015fea} was a factor of $\sim$$2.4$ larger than our current mask, which reduced the available ROI in the inner few degrees of the Galaxy. 

\subsection{Signal Injection Tests on Data}

\begin{figure*}[htb]
\leavevmode
\begin{center}
\includegraphics[width=1.0\textwidth]{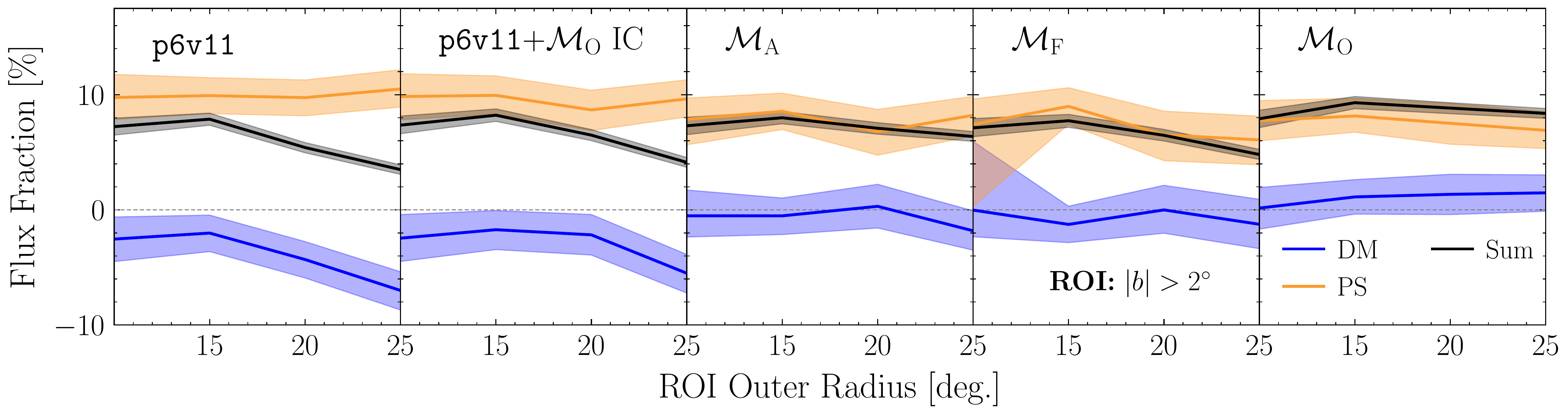}
\end{center}
\caption{As in Fig.~\ref{fig:NPTF-harm-data}, except for the analysis where we change the outer radius of our ROI away from the fiducial value $r_{\max} = 25^\circ$ to the indicated value.  Note that even though the analyses are computed in different ROIs, the flux fractions are still computed relative to our fiducial ROI to make it easier to compare results between different ROIs.  The results for Models O, F, and A indicate no evidence for a non-zero DM flux fraction across the full ensemble of ROIs.  On the other hand, the \texttt{p6v11} results show evidence for over-subtraction at large $r_{\rm max}$, but this is partially mitigated by going to smaller ROIs.  Note that the harmonic marginalization is not performed here. }
\label{fig:ROI-results}
\end{figure*} 

As we have shown, the \texttt{p6v11} diffuse model suffers from over-subtraction and one manifestation of this is that the normalization of the Poissonian GCE template is driven to negative values in the NPTF, if allowed by the priors.  Model O, on the other hand, does not appear to suffer from this systematic bias.  Another way of understanding these results is in the context of the signal-injection tests presented by Leane~and~Slatyer~\cite{Leane:2019uhc}.  In that work, the authors injected a synthetic DM signal on top of the real {\it Fermi} data and analyzed the hybrid data for evidence of GCE-correlated PSs and GCE Poissonian emission using the NPTF.  The posterior recovered for the GCE Poissonian emission did not include the simulated value, when using the \texttt{p6v11} diffuse model, and this was taken as evidence that the NPTF is not a trustworthy diagnostic to distinguish emission from GCE-correlated PSs and DM.  However, we now understand that this result was driven by mismodeling in the \texttt{p6v11} diffuse model.\footnote{We note that in Ref.~\cite{Leane:2019uhc} the authors also demonstrated that 2\% injected GCE flux fractions are not recovered in Model~A and~F.
This is consistent with the mild over-subtraction we see for these diffuse models in Fig.~\ref{fig:NPTF-harm-data}.
Although we only explicitly demonstrate harmonic marginalization resolves the signal injection issue for \texttt{p6v11} in this subsection, in Fig.~\ref{fig:NPTF-harm-data} we can see that for higher harmonics the over-subtraction is relieved, and therefore we expect a similar conclusion to hold for Model~A and~F as well.}

In the left panel of Fig.~\ref{fig:NPTF-harm-data-sig-inj}, we repeat the signal injection test on {\it Fermi} data and show the recovered Poissonian GCE flux fraction on the $y$-axis.  We stress that the NPTF analysis includes both a Poissonian GCE template (with strictly non-negative prior) and a GCE-correlated PS template.  When the \texttt{p6v11} diffuse model is used in the analysis, we do not recover the expected result, which is indicated by the dashed diagonal blue line.  This is because the DM normalization wants to be negative and so a sufficient amount of flux must be injected (around 7\% in flux fraction as shown in Fig.~\ref{fig:NPTF-harm-data}) before the recovered flux fraction begins to rise above zero.  On the other hand, the results found when using Model O (red) are more consistent with expectations.  

The spherical-harmonic marginalization procedure is able to partially mitigate the over-subtraction for \texttt{p6v11}.  One sign of this is that, as shown in the right panel of Fig.~\ref{fig:NPTF-harm-data-sig-inj}, after applying the harmonic marginalization procedure to the diffuse models through $\ell = m = 2$, both the Model~O and \texttt{p6v11} models behave better under the injected signal test.  In particular, now when using the (improved) \texttt{p6v11} diffuse model, we find agreement between the injected and recovered flux fractions for a synthetic DM signal.

\section{The effect of the ROI on diffuse mismodeling}
\label{sec:ROI}

This section presents an alternate strategy for mitigating the over-subtraction that was pointed out for the \texttt{p6v11} diffuse model.
The \texttt{p6v11} model was developed to search for individual resolved PSs in small patches of the sky and care must therefore be taken when using the model to study large-scale features, like the GCE.  In particular, while the diffuse models may correctly describe the small-scale structure in the data, they may fail to capture large-scale variations, potentially biasing the fit results.  We have already demonstrated that harmonic marginalization can mitigate such issues on the data.  Here, we compare these results to those obtained using an alternate, less sophisticated strategy: to simply reduce the size of the ROI.  The disadvantage of this approach is that we reduce the overall photon count by making the ROI smaller, thereby losing sensitivity. However, it still provides a useful counterpoint to the harmonic marginalization results presented earlier.  

As a reminder, the fiducial ROI was defined as $|b| \geq 2^\circ$ and $r \leq r_\text{max}$, where $r_\text{max} = 25^\circ$. We now progressively reduce the outer radius of the ROI, $r_{\rm max}$, from 25$^\circ$ to $10$$^\circ$.   Note that we continue to use the 3FGL source mask throughout.  In Fig.~\ref{fig:ROI-results}, we show the flux fractions recovered for the GCE DM and PS templates as a function of $r_{\rm max}$ from an analysis of the actual {\it Fermi} data.  Note that the different panels correspond to different diffuse templates.

\begin{figure}[t]
\leavevmode
\begin{center}
\includegraphics[width=0.48\textwidth]{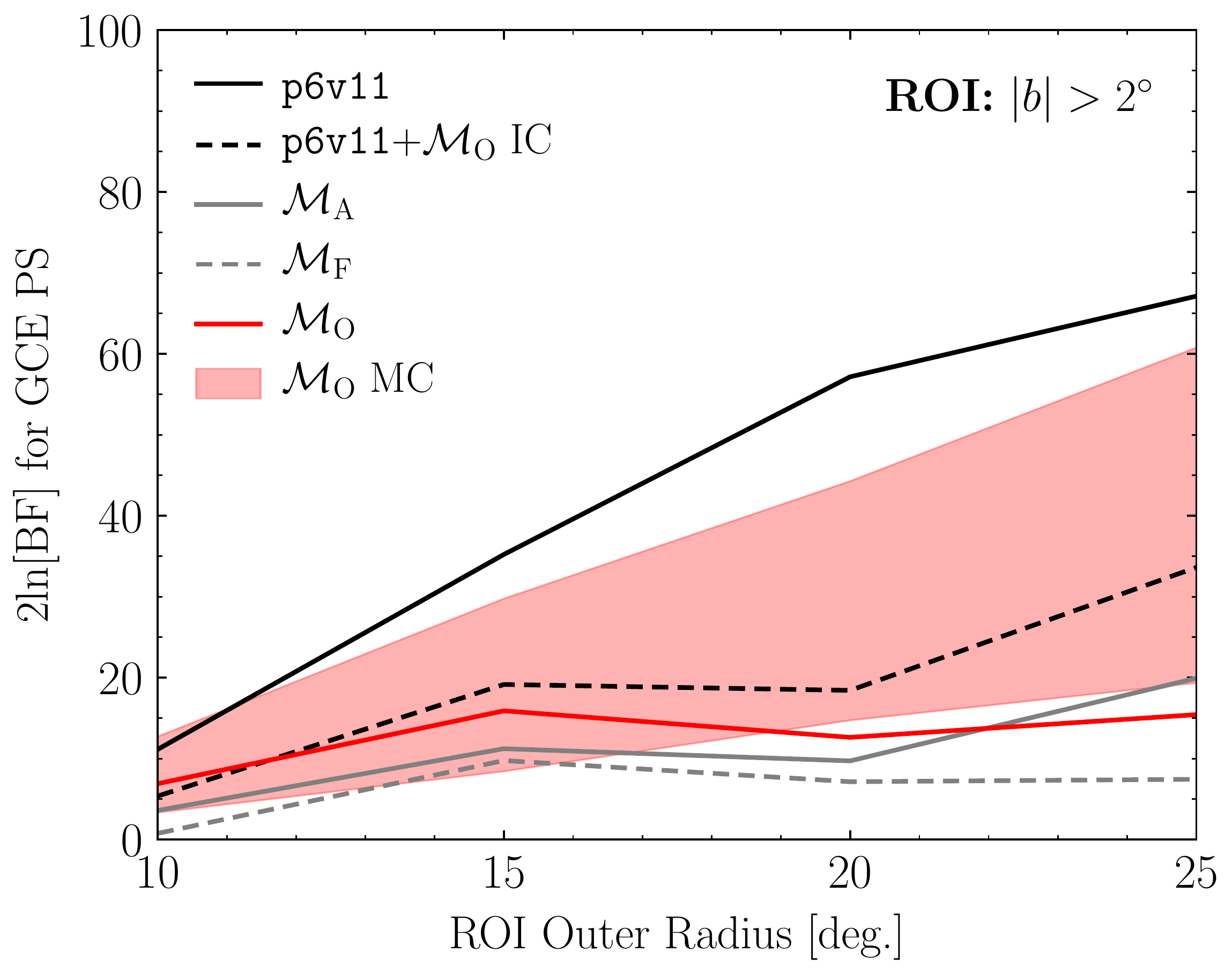}
\end{center}
\caption{The evidence in favor of GCE PSs over DM for the analyses shown in Fig.~\ref{fig:ROI-results}, except in this case the DM flux fraction is restricted to be non-negative.  All models show evidence for GCE PSs over DM for $r_{\rm max} \gtrsim 15^\circ$.  In the case of \texttt{p6v11} only, we see that at large $r_{\rm max}$ over-subtraction leads to inflated evidence in favor of the model with PSs.  Note that the harmonic marginalization is not performed here. The red band represents the MC expectation, depicted as the region between the 16 and 84 percentiles from eleven simulations. The simulation is constructed from the best fit Model~O template on data, using an ROI with $r_{\rm max} = 15^{\circ}$.
}
\label{fig:ROI-BF-results}
\end{figure}

With the Model~O templates, we find that the flux fractions recovered for DM and PSs remain relatively consistent for the different $r_{\rm max}$.  On the other hand, when we use the \texttt{p6v11} template, the results are seen to vary wildly as a function of $r_{\rm max}$.  At large $r_{\rm max}$, over-subtraction is a serious issue and the DM template normalization is driven to significantly negative values.  On the other hand, the over-subtraction is partially mitigated by going to smaller $r_\text{max}$, as would be expected.  This trend is also reflected in the Bayes factors, as seen in Fig.~\ref{fig:ROI-BF-results}.

In the Model O case, the Bayes factor in preference for PSs remains relatively constant with $r_{\rm max}$ until $r_{\rm max} \lesssim 15^\circ$. The MC expectation is shown as the red band for Model~O, and in all cases is constructed from eleven simulations of the best fit parameters obtained in an ROI with $r_{\rm max} = 15^{\circ}$, at which radius we also see consistency with the data.
For \texttt{p6v11}, the Bayes factor falls substantially with decreasing $r_{\rm max}$, which suggests that the large Bayes factors seen at high $r_{\rm max}$ are inflated by over-subtraction.  These results, just like the harmonic marginalization results presented previously, suggest that \emph{(i)}~the \texttt{p6v11} diffuse model may substantially bias searches for dim PSs in the Inner Galaxy, and \emph{(ii)}~even after mitigating diffuse mismodeling, in this case by reducing the ROI and changing diffuse models, the evidence in favor of PSs over DM remains robust.

\section{Discussion}

In this paper, we discuss how the evidence in favor of PSs for the GCE is influenced by mismodeling of the Galactic diffuse emission.  The GCE is a subdominant component of the gamma-ray flux observed by {\it Fermi} in the Inner Galaxy, so mismodeling the Milky Way foreground emission---which contributes the bulk of the flux---can bias the properties recovered for the GCE.  The effects of diffuse mismodeling on the NPTF procedure were first explored in Ref.~\cite{Lee:2015fea} but recently revisited in Ref.~\cite{Leane:2019uhc}.  In this work we present a dedicated study of how diffuse mismodeling affects the evidence for unresolved PSs in the Inner Galaxy, which we have defined primarily as $|b| > 2^\circ$ and $r < 25^\circ$.

We consider four different foreground models: \texttt{p6v11}, as well as Models~A,~F, and~O.  The \texttt{p6v11} diffuse model has been  used as a standard benchmark in GCE studies and Models~A and~F have also been commonly used in the literature.  Model~O is a new state-of-the-art diffuse emission model, based on those in Refs.~\cite{Macias:2016nev,Macias:2019omb}, that we construct for this analysis.\footnote{This diffuse model is available \href{https://github.com/nickrodd/FermiDiffuse-ModelO}{here}.} Of these, \texttt{p6v11} provides by far the worst fit to the {\it Fermi} data in the Inner Galaxy, while Model~O provides the best.  We show that \texttt{p6v11} results in serious over-subtraction when performing template fits on the data with 3FGL sources masked.  (The same applies for Model~F, although to a lesser degree.)  This biases the properties of the GCE recovered by both Poissonian and non-Poissonian template fits.

In addition to exploring the effects of Model~O, we also introduce a new statistical procedure, called spherical-harmonic marginalization, to further mitigate  mismodeling effects on large angular scales.  We find that, when applying this procedure, the results for all four diffuse models often converge towards large $\ell_\text{max}$, the maximum spherical-harmonic number that is marginalized over, suggesting that they all yield consistent results once large-scale mismodeling effects are minimized. In particular, the over-subtraction issues that are particularly striking for \texttt{p6v11} and Model~F are resolved.  This gives us confidence that the spherical-harmonic marginalization procedure successfully tempers issues associated with diffuse mismodeling on large angular scales.  

From our close study of the effects of diffuse mismodeling on the NPTF, we reach two primary conclusions:

\begin{itemize}
\item The evidence in favor of PSs over DM for the GCE is robust, at least to the extent that we can test for diffuse mismodeling and assuming an NFW distribution for both the PS population and the DM.  The original NPTF study~\cite{Lee:2015fea} primarily used \texttt{p6v11}, though fourteen other diffuse models were also explored in the Appendix.  At the time, the evidence for PSs was observed to be fairly consistent across all models and was $\sim$$10^4$ for the 3FGL-masked \texttt{p6v11} analysis, though that 3FGL mask was significantly larger than the one used here and masked much of the inner regions of the Galaxy.  In this work, we find that the preference in favor of PSs is $\sim$$ 10^{3.4}$ for Model~O, and remains essentially constant after applying the harmonic marginalization procedure.  The \texttt{p6v11} Bayes factor starts off much higher---likely due to residuals from the diffuse mismodeling that masquerade as PSs---but approaches the Model~O Bayes factor once large-scale modes are given additional freedom through the harmonic-marginalization procedure.
\item The signal injection tests reported by Leane~and~Slatyer~\cite{Leane:2019uhc} are explained by their choice of diffuse model.  They found that an artificial DM signal that was injected into the {\it Fermi} data was not properly recovered by the NPTF.  The authors used this result to question the robustness of PS explanation for the GCE.  We clearly demonstrate that the result of the signal injection test is due to over-subtraction resulting from the choice of diffuse models.  When, for example, Model~O is used, the signal injection test works as expected, with the correct DM flux recovered.  Therefore, we conclude that the apparent inconsistency pointed out by Leane~and~Slatyer~\cite{Leane:2019uhc} has an understandable origin in terms of diffuse mismodeling. 
\end{itemize}

We end now with some important cautionary notes on how to interpret the results presented in this work, as well as other GCE studies:

\begin{itemize}
    \item Diffuse mismodeling is currently the most important  systematic uncertainty in any study of the GCE.  Over/under-subtraction of the foregrounds can easily affect the characterization of the flux, morphology, and energy spectrum of the Excess.  We argue in this work that Model~O provides an improved fit to the data, compared to \texttt{p6v11} and Models~A and~F, three standard models used in the literature. However, we do not claim that Model~O is necessarily the final or best answer. (See App.~\ref{app:qual} where we show that Model~O is not modeling the data at the level of statistical noise.)  Continued improvements to modeling the Milky Way's diffuse emission is the singularly most important effort needed to characterize the nature of the GCE. 
    \item The choice of templates used in any {\it Fermi} study is another relevant uncertainty.  This is true for either a standard (Poissonian) template analysis or the NPTF.  In this paper, we focus on comparing the NFW~DM template against an NFW~PS template.  We do not consider variations to the DM or PS templates, as our focus is on improving the diffuse emission modeling.  If there is an emission component in the data that we do not model with a template, or if the assumption of the NFW distribution is incorrect, it could potentially affect the Bayes factor preference for PSs.  This was pointed out in the original NPTF study~\cite{Lee:2015fea}, and more recently in the context of gas clumps tracing the {\it Fermi} bubbles in Ref.~\cite{Leane:2019uhc}.  This is a standard challenge of  any template analysis, as highlighted by recent evidence that the excess itself may be better correlated with stellar overdensities~\cite{Macias:2016nev,Macias:2019omb,Bartels:2017vsx}.
    \item The NPTF is agnostic to the nature of the source population.  When we say that the evidence for unresolved PSs remains robust, we do not make any claims as to the nature of those sources.  Millisecond pulsars are a candidate, as their energy spectrum is roughly consistent with that of the GCE.  However, it may also be that the evidence for PSs arises from  small-scale structures in the foreground model that are not properly captured by the diffuse template.  This was a concern in the original NPTF analyses~\cite{Lee:2015fea} and it remains true today.  Additionally, it need not be one unique population of sources that contributes to the evidence for PSs.
    \item Our results do not exclude the possibility that some fraction of the GCE is DM and the other fraction is PSs, as already cautioned  in Ref.~\cite{Lee:2015fea}.  The ability of the NPTF to recover the true fraction of DM and PSs in such mixed scenarios was studied in our companion paper~\cite{Chang:2019ars}.  The challenge arises from the fundamental degeneracy between faint PSs and smooth DM emission.
    \item The spherical-harmonic marginalization procedure that we introduce here is not well setup to deal with  small-scale issues in the diffuse model because the $\ell \leq 4$ harmonic modes that are marginalized over modulate large angular scales.  Thus, one possibility is that the bulk of the GCE arises from DM annihilation and additionally there are small-scale mismodeling effects with the diffuse model that cause the NPTF to infer that the whole GCE arises from PSs.  While there is no evidence at present that points to this conclusion, we also cannot disprove it entirely.
    
    This issue may be particularly relevant, however, when considering that the gas and dust maps that go into making the diffuse models are smoothed at angular scales comparable to but larger than the {\it Fermi} PSF (for example, for Model~O, the maps are smoothed at an angular resolution $\sim$$0.5^\circ$).  Recently, however, a full-sky H$_{\rm I}$ map (called \texttt{HI4PI}) was constructed at an angular resolution $\sim$$0.25^\circ$ using the EBHIS and GASS surveys~\cite{2016A&A...594A.116H}.  In App.~\ref{app:HI}, we construct an additional diffuse emission template based off of the \texttt{HI4PI} map.  We find that while the inclusion of this template substantially improves the fit to the {\it Fermi} data, it does not qualitatively effect the evidence for GCE-correlated PSs.  In fact, the evidence in favor of GCE-correlated PSs increases slightly with this inclusion of the \texttt{HI4PI} map.  Still, this does not rule out the possibility that with future higher-resolution diffuse foreground templates the evidence in favor of PSs will be reduced.
\end{itemize}

Given the significant effects that the diffuse foreground model has on reconstructing the properties of the GCE, care must be taken when making claims of a DM origin for the excess.  The current pieces of evidence indicate that there are standard astrophysical contributions to the GCE---whether these be actual source populations like millisecond pulsars or emission from mismodeled components of the diffuse foreground.  At this stage, it is not possible to know whether these effects can explain the entirety of the GCE.  It may very well be that DM is present and sitting below these more standard contributions.  Before any such claim can be made irrefutable, however, these astrophysical contributions must be carefully studied and robustly characterized.   

\section{Acknowledgements}
\label{sec:acknowledgements}
We thank P.~Fox, D.~Hooper, R.~Leane, S.~McDermott, S.~Murgia, K.~Perez, T.~Slatyer, T.~Tait, K.~Van~Tilburg, N.~Weiner, and C.~Weniger for useful conversations. LJC is supported by a Paul \& Daisy Soros Fellowship and an NSF Graduate Research Fellowship under Grant Number DGE-1656466. ML is supported by the DOE under Award Number DESC0007968 and the Cottrell Scholar Program through the Research Corporation for Science Advancement. SM is supported by the NSF CAREER grant PHY-1554858, NSF grants PHY-1620727 and PHY-1915409, and the Simons Foundation. NLR is supported by the Miller Institute for Basic Research in Science at the University of California, Berkeley. MB and BRS are supported by the DOE Early Career Grant DESC0019225. OM acknowledges support by JSPS KAKENHI Grant Numbers JP17H04836, JP18H04340, JP18H04578 and by World Premier International Research Center Initiative (WPI Initiative), MEXT, Japan. This work was performed in part at the Aspen Center for Physics, which is supported by National Science Foundation grant PHY-1607611. 
 This work was supported through computational resources and services provided by Advanced Research Computing at the University  of  Michigan,  Ann  Arbor. This work made use of resources provided by the National Energy Research Scientific Computing Center, a U.S. Department of Energy Office of Science User Facility supported by Contract No. DE-AC02-05CH11231.
 
\newpage

\appendix 

\section{4FGL point source mask}
\label{app:4FGL}

\begin{figure*}[tbh]
\leavevmode
\begin{center}
\includegraphics[width=0.47\textwidth]{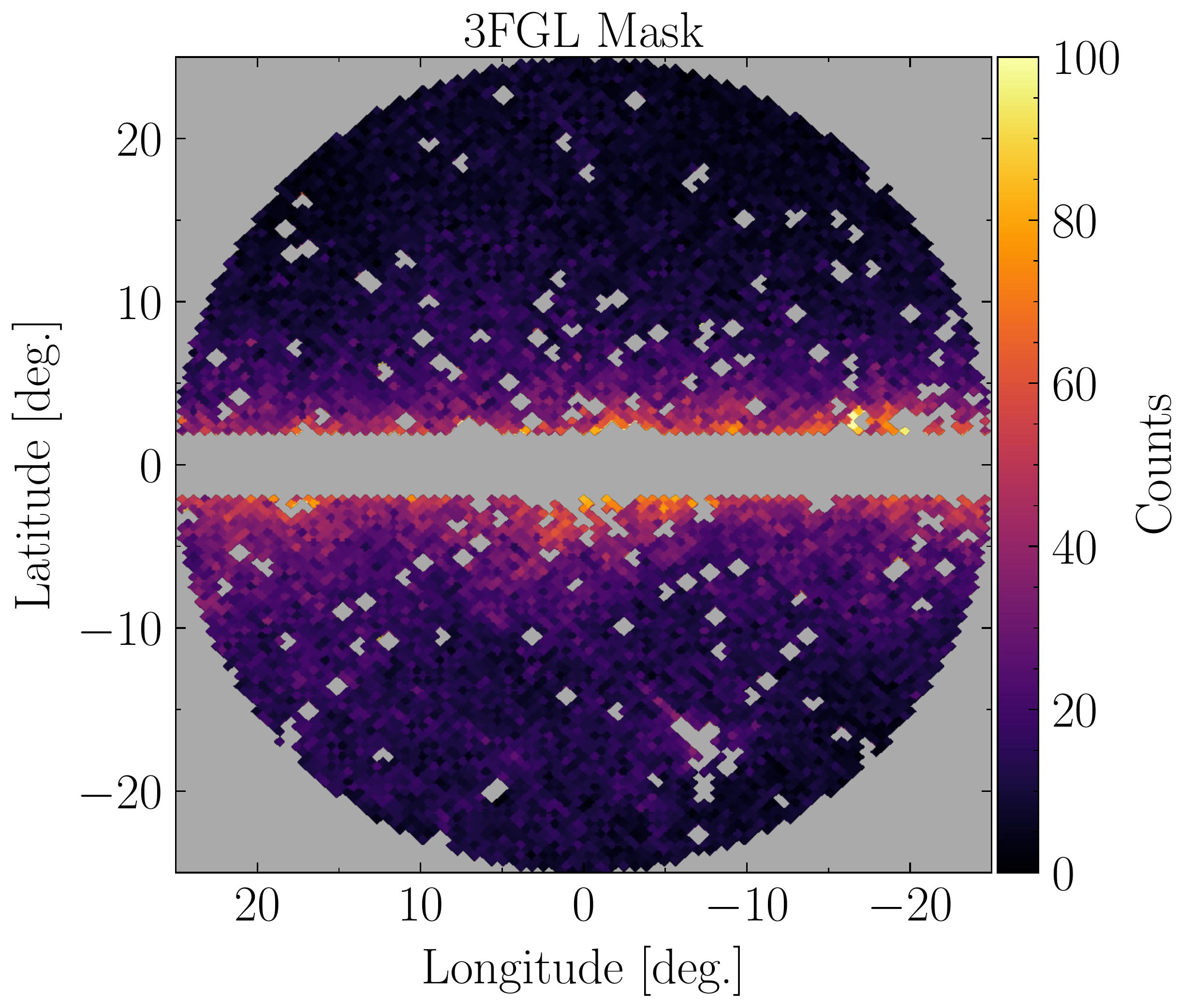} 
\hspace{0.5cm}
\includegraphics[width=0.47\textwidth]{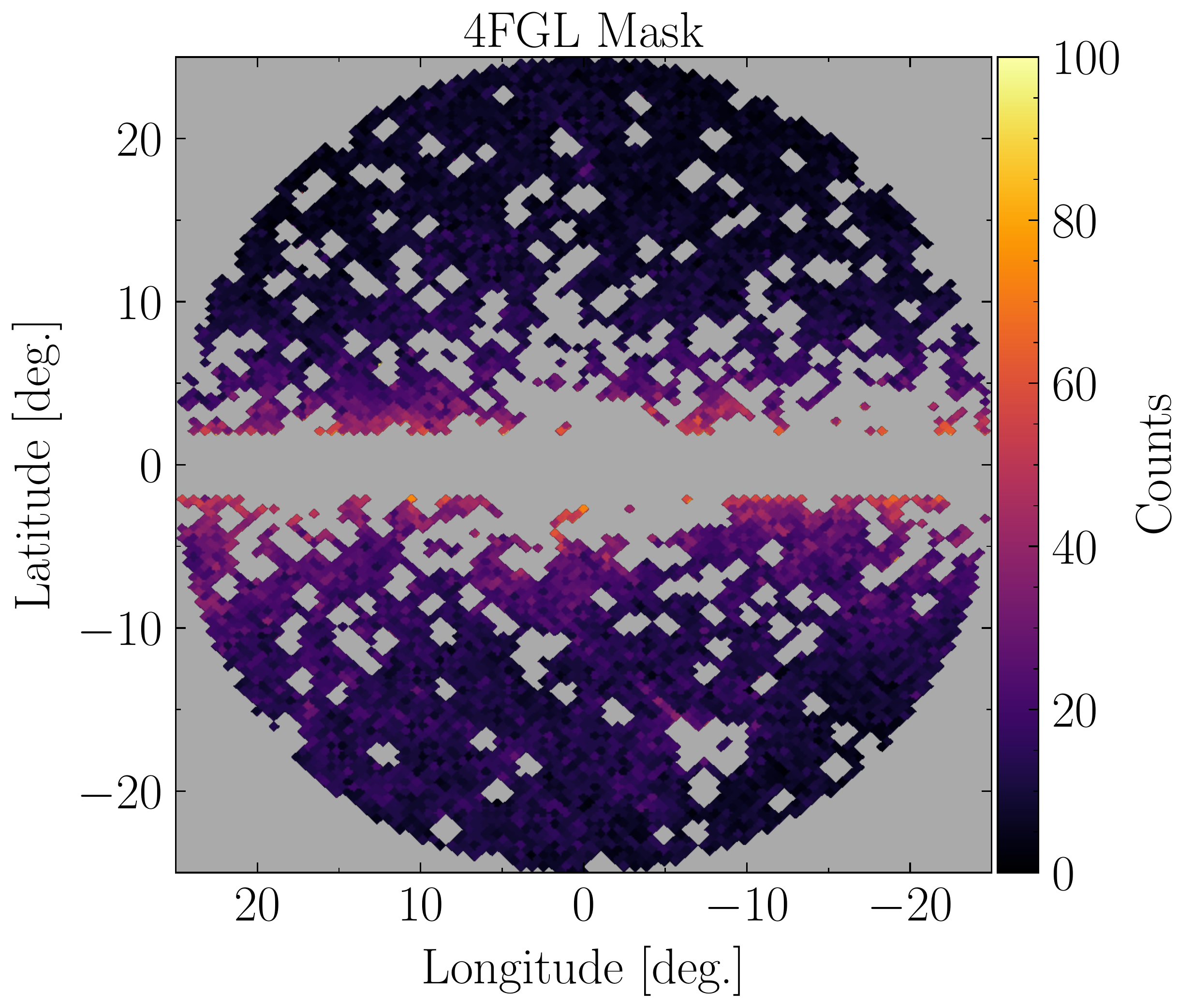} 
\end{center}
\caption{(Left) The {\it Fermi} data as observed in our default ROI: $|b|>2^{\circ}$, $r<25^{\circ}$, and 3FGL sources masked, where masked pixels are colored gray.
(Right) As on the left, but now using the 4FGL source catalog, which removes a significant fraction of the ROI.}
\label{roi:compare}
\end{figure*}

In this section, we explore the effect of replacing the 3FGL PS mask used in our default analysis with the newer 4FGL PS mask~\cite{Fermi-LAT:2019yla}.  The 4FGL catalog includes almost 70\% more sources than the older 3FGL catalog.  It is thus an interesting question to ask how our results change when these additional sources are masked.  We construct a PS mask from the 4FGL catalog using the same procedure that we apply to make the 3FGL mask. In particular, all sources are masked at the 95\% PSF containment radius as defined in our lowest energy bin, which evaluates to $0.47^{\circ}$.

One significant difference between using the 3FGL and 4FGL PS masks is that using the 4FGL mask severely reduces the area near the GC in the fiducial ROI.  This is illustrated in Fig.~\ref{roi:compare}, where we compare the fiducial ROI using the 3FGL mask to the ROI used with the 4FGL mask. If we take a region defined by $|b|>2^{\circ}$ and $r<25^{\circ}$, masking 3FGL sources covers 7\% of the region, whereas the 4FGL covers 28\%. This additional masking is preferentially towards the Galactic Center, where the NPTF draws on much of its power---see Fig.~\ref{fig:TS-Maps}. If we considered a ROI of $|b|>2^{\circ}$ and $r<10^{\circ}$, the 3FGL covers 13\% of this region, whereas the 4FGL masks more than 50\%.

The first question to ask before analyzing the real data, is what do we expect for the reduction of evidence in favor of PSs due to the reduction in ROI size.  Towards that end, we analyze the same MC datasets generated in Sec.~\ref{sec:harm-toy} for a PS GCE but with the 4FGL PS mask instead of the 3FGL PS mask. Recall, the MC from Sec.~\ref{sec:harm-toy} was generated from the best-fit parameters of a Model~O analysis within $r<15^{\circ}$. Using this, we found in our default $r<25^{\circ}$ and 3FGL-masked ROI, a median Bayes factor in favor of PSs of $\sim$$10^{10}$, and with a 68\% expected range from $10^4$--$10^{13}$, when analyzing the PS MC with the true diffuse model (Model O).  The reduced ROI from the 4FGL mask leads to noticeably smaller Bayes factors: the median is now $10^5$, with a 68\% range of $10^{2.6}$--$10^7$. This is interesting because in the MC case, the PSs are distributed randomly throughout the ROI and are therefore not correlated with the mask.  This means that the reduction in Bayes factor in the MC is being driven by the smaller ROI size.  On data, both the 3FGL mask---and to a greater extent, the 4FGL mask---will by definition preferentially mask actual PSs. Thus, we expect the Bayes factors to be even smaller on data than the expected range in MC, where the PS locations are not correlated with the mask.

Indeed, when we analyze the actual {\it Fermi} data with the 4FGL mask, we recover a Bayes factor of 2.6 (in the 3FGL case, we find $10^{3.4}$), which implies negligible evidence for GCE PSs, and is in agreement with the recent findings of Ref.~\cite{Zhong:2019ycb}.  However, it is important to properly interpret this statement.  Finding negligible evidence for GCE PSs with the 4FGL mask does not mean that the GCE is not arising from PSs.  Rather, it can also be due to the fact that the statistics are not sufficient to definitely say whether the GCE is arising from PSs.  There are two reasons for this.  First, the ROI is simply not large enough, as we understood from the MC tests, to make a strong statement about the PS origin of the GCE.  And secondly, the 4FGL mask likely removes many of the brighter members of the GCE PS population, if such a population exists, which were not already removed by the 3FGL mask.  

We can try to estimate how many GCE-correlated PSs would be removed in going from the 3FGL to 4FGL mask by approximating the sensitivity differences between the two catalogs.  Doing this exercise in a principled way would require modeling the dependence of the 4FGL detection sensitivity on the spatial location of the sources and the source spectral properties, as was performed in \emph{e.g.}, Refs.~\cite{Fermi-LAT:2017yoi,Bartels:2017xba}, which is beyond the scope of this paper.  Still, as a rough estimate, we may do the following.  In our fiducial ROI and energy range, we expect that the approximate 3FGL detection threshold is around $3 \times 10^{-10}$~counts/cm$^2$/s (see, {\it e.g.}, Ref.~\cite{Lee:2015fea}).  The 4FGL catalog was constructed with approximately twice the exposure time as the 3FGL catalog.  Since the PS searches are background dominated, this implies that the flux sensitivity should approximately increase by an amount $\sim$$\sqrt{2}$, so that sources with fluxes $\sim$$2 \times 10^{-10}$~counts/cm$^2$/s and above would be detectable.  Of course, in reality, this flux sensitivity depends on where the source happens to be located (sources closer to the GC are harder to detect than those further away) and also on the spectral characteristics of the source. 

Still, it is interesting to ask how the flux fraction associated with the GCE would change if we were to mask all sources between $2 \times 10^{-10}$ counts/cm$^2$/s and $3 \times 10^{-10}$~counts/cm$^2$/s.  By integrating the flux-weighted source-count distribution given in Fig.~\ref{fig:NPTF-MC-dNdF}, we find that masking the additional sources should reduce the normalization of the GCE by $\sim$10\% in the case where the GCE arises from spherical PSs (a consistent result is obtain using the source-count distribution given in Fig.~\ref{fig:unmasked_lumi} below). To calculate the number of GCE-correlated PSs that this corresponds to, we may again integrate our source-count distribution (this time not flux weighted) between $(2-3) \times 10^{-10}$ counts/cm$^2$/s, and we find that this 10\% decrease in flux fraction is arising from only a handful of PSs ($\sim$4).  In practice, there are many more than 4 additional PSs within our fiducial ROI between the 4FGL and 3FGL catalog, but it is important to remember that this is a rough estimate  that does not account for the additional disk-correlated and isotropic sources that appear in 4FGL.  In terms of the flux fraction, interestingly we do observe that the GCE flux fraction (summed between GCE-correlated PSs and GCE-correlated Poissonian emission) decreases by $\sim$9.2\% when going from the 3FGL mask to the 4FGL mask.  However, this latter result should also be interpreted with care, since when going to the 4FGL mask, we also considerably shrink the region, and the normalization of the GCE is known to depend sensitively on the ROI, as illustrated in {\it e.g.}, Sec.~\ref{sec:poiss}.

\section{3FGL-Unmasked NPTF results}
\label{app:unmasked}

Throughout the main body of this work, we always include a 3FGL PS mask.  The reason for this is two-fold.  First, this allows us to more directly compare to central results in Ref.~\cite{Leane:2019uhc}.  Second, many of the 3FGL sources are high-flux sources that do not contribute to the GCE but do make the evaluation of the NPTF likelihood computationally more costly.  Related to the second point is the worry that when not masking known PSs, we may become more sensitive to {\it e.g.}, the exact form of the disk PS template included.  At the least, we open ourselves up to the additional source of systematic uncertainty which is disk-correlated or isotropically-distributed PSs masquerading as GCE-correlated PSs.  Still, in this Appendix, we study the properties of the GCE when analyzed using Model~O and without the 3FGL PS mask.

\begin{figure}[t]
\leavevmode
\begin{center}
\includegraphics[width=0.5\textwidth]{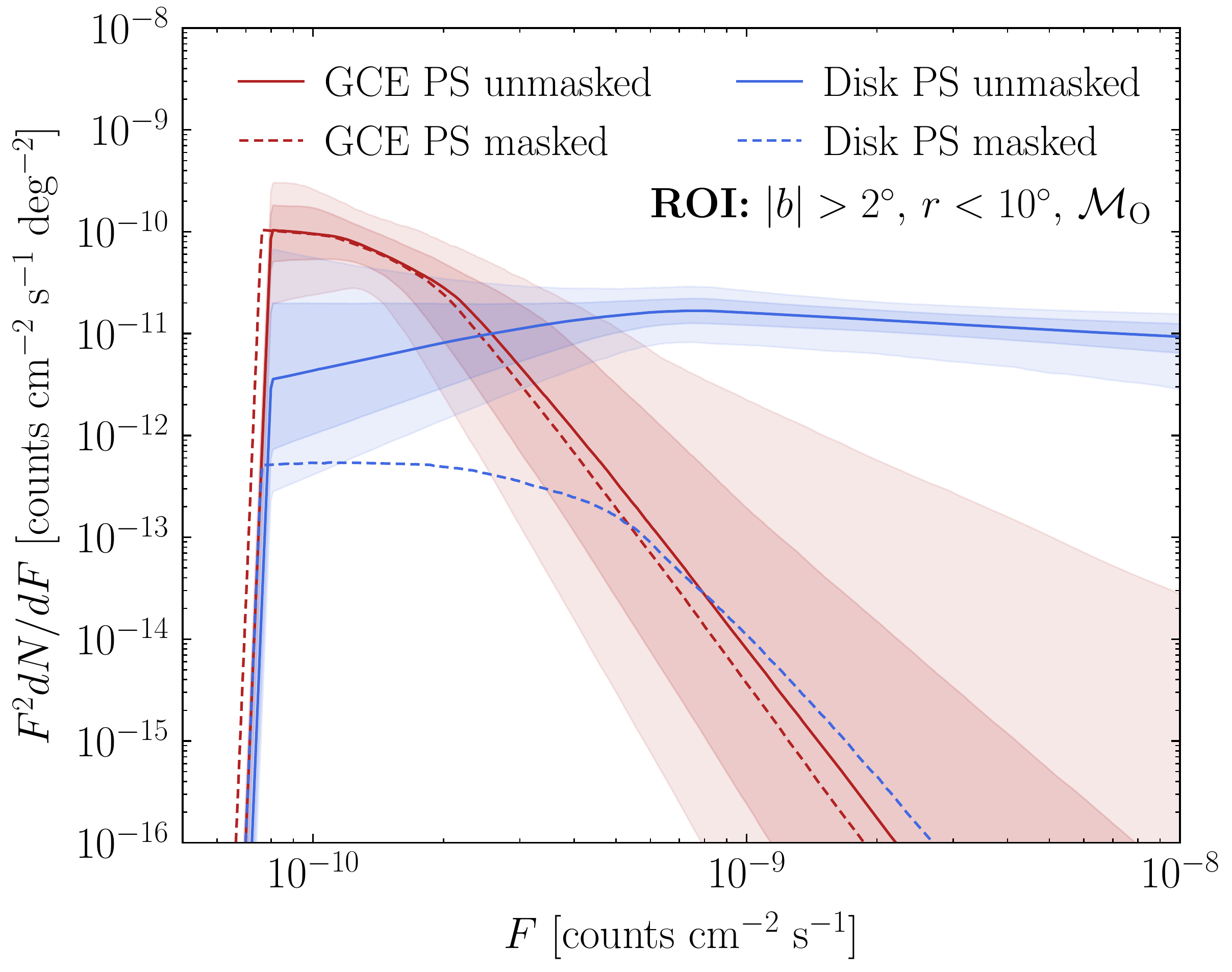}
\end{center}
\caption{The source-count distribution for the GCE-correlated and disk-correlated PSs (as in Fig.~\ref{fig:NPTF-MC-dNdF}) from a 3FGL-unmasked analysis in the ROI defined by $|b| > 2^\circ$ and $r < 10^\circ$.  This analysis uses the Model O diffuse model.  The GCE-correlated source-count distribution is consistent with that found in the unmasked analysis, but in this case, the disk source-count distribution extends to high fluxes in order to explain the 3FGL sources.}
\label{fig:unmasked_lumi}
\end{figure}

In this case, it is interesting to focus on the inner 10$^\circ$ around the GC in order to minimize diffuse mismodeling.  In the masked case, this produced a small Bayes factor in preference for spherical PSs with Model~O ($2 \ln [{\rm BF}] \sim 7$), but in the unmasked case, more of the Inner Galaxy region is available and so if the GCE is truly arising from PSs, we would expect this number to go up.  Indeed, in the unmasked analysis, we find a Bayes factor in preference for spherical PSs of $2 \ln [{\rm BF}] \sim 12$.  The flux fraction from GCE-correlated PSs, normalized to our fiducial ROI, is consistent between the masked and unmasked analyses ($7.7 \pm 1.8$ in the masked analysis and $7.7 \pm 1.7$ in the unmasked analysis).  Note, however, that the flux fraction in disk-correlated PSs increases dramatically, as expected, when we unmask the 3FGL sources ($0.1_{-0.1}^{+0.9}$ in the masked analysis and $10.7_{-3.5}^{+6.7}$ in the unmasked analysis).  In both cases, the DM flux fraction is consistent with zero.  

The source-count distributions recovered for the GCE-correlated and disk-correlated PSs are shown in Fig.~\ref{fig:unmasked_lumi}.  Note that the GCE-correlated PS source-count distribution is consistent with that found in the masked analysis.  However, without the 3FGL mask, the disk source-count distribution has support at  large fluxes, as expected.

\section{Quality of Fit of Diffuse Models}
\label{app:qual}
\begin{figure}[t]
\leavevmode
\begin{center}
\includegraphics[width=0.5\textwidth]{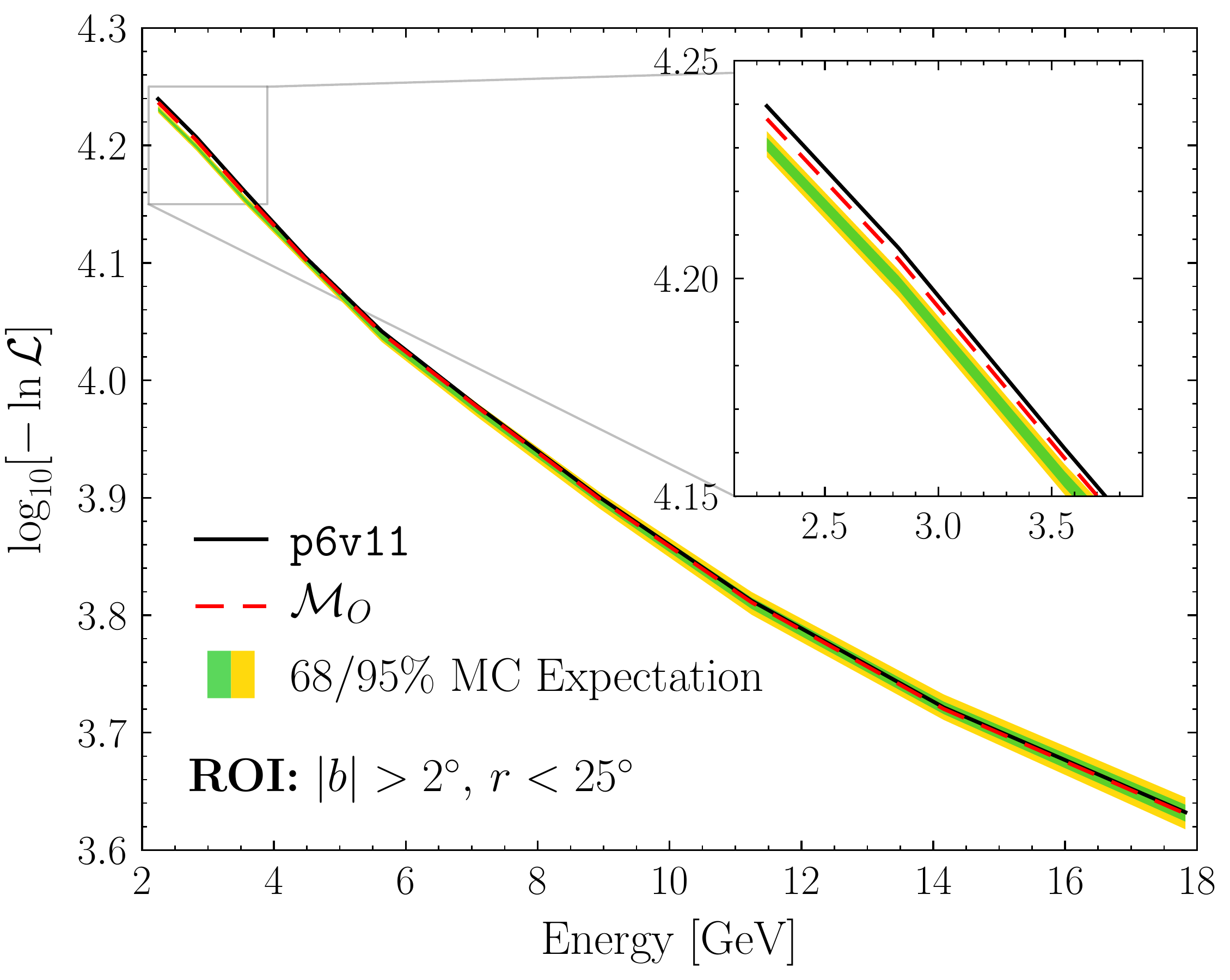}
\end{center}
\caption{The likelihood values (plotted as $\log_{\rm 10} [-\ln{ {\mathcal L}}]$) for the Poissonian analyses whose spectra are shown in Fig.~\ref{fig:GCE-spec} (Model O and \texttt{p6v11} only).  To interpret the absolute goodness of fit of the diffuse models, we compare these log-likelihood values to the same quantities computed from MC under the hypothesis that the data is a Poisson draw of the best-fit templates for the Model O analysis.  This leads to the expectations shown in green and yellow at 68\% and 95\% confidence, respectively.  While Model O is a substantially better fit than \texttt{p6v11}, at energies below $\sim$4 GeV, we also see that Model O does not describe the data to the level of Poisson noise. }
\label{fig:FitQuality}
\end{figure}

In this section, we consider the absolute goodness-of-fit of the diffuse models used in this work to the {\it Fermi} data. We have already seen throughout the main body that Model~O is a better description of the data relative to \texttt{p6v11}.  Here, we consider the quality of fit in absolute terms.  To do so, we use the Poissonian analysis in the default ten logarithmically-spaced energy bins between 2 and 20 GeV used in this work.  Naively, we could determine the quality of fit by, for example, computing the $\chi^2$ per degree of freedom. Yet there are pixels in our dataset with as few as 0 or 1 photons in them, where the fit will not be $\chi^2$ distributed.  Therefore, we perform the following procedure.  First, in a given energy bin, we fit the Poissonian model (including the diffuse emission model, isotropic emission, {\it Fermi} bubble emission, GCE DM, and 3FGL emission from outside of the 3FGL mask) to the {\it Fermi} data.  From this fit, we obtain a value for $\ln{ {\mathcal L}}$, where ${\mathcal L}$ is the Poisson likelihood.  We perform these fits in our ficucial ROI ($|b| > 2^\circ$, and $r < 25^\circ$) with 3FGL sources masked.  The results of this analysis for both the \texttt{p6v11} and Model O diffuse models are shown in Fig.~\ref{fig:FitQuality}.  Note that smaller values of $\log_{\rm 10} [-\ln{ {\mathcal L}}]$ indicate larger likelihood values and thus better fits to the data.  As expected, Model O outperforms \texttt{p6v11} across the whole energy range. 

To interpret the $\log_{\rm 10} [-\ln{ {\mathcal L}}]$ values, it is useful to understand their expectations under the scenario where the {\it Fermi} data is a Poissonian draw of the best-fit template sum from the analysis on the real data.  For this exercise, we generate one thousand MC realizations of simulated datasets constructed from the best-fit model when using the Model O diffuse templates (the band obtained if we instead used \texttt{p6v11} is similar). For each MC, we  compute the $\ln{ {\mathcal L}}$ in each energy bin as compared with the model it was drawn from and then look at the distribution of values.  The 68\% and 95\% expectations for $\log_{\rm 10} [-\ln{ {\mathcal L}}]$ are indicated in Fig.~\ref{fig:FitQuality}. At high energies ($E \gtrsim 4$ GeV), Model O describes the data to the level of Poisson noise.  However, at lower energies, even though Model O is a considerable improvement on \texttt{p6v11}, there remains a systematic discrepancy between the Poisson noise expectation and the likelihood values observed on real data. We therefore conclude that diffuse mismodeling likely dominates over the statistical uncertainties in analyses in this ROI, for all diffuse models employed in this work. More quantitatively, the difference between Model O and the Poisson noise expectation in the lowest energy bin, as measured by twice the log-likelihood ($2 \times \Delta \ln {\mathcal L}$) between the value observed on data and the average value from MC is 465.

\section{Effect of High-resolution H$_{\rm I}$ Gas Template}
\label{app:HI}
In this Appendix, we construct a high angular resolution gas template from the full-sky H$_{\rm I}$ map \texttt{HI4PI} produced in Ref.~\cite{2016A&A...594A.116H}.  We begin with the \texttt{HEALPix} \texttt{HI4PI} map produced in Ref.~\cite{2016A&A...594A.116H}, and then we pass this map through the instrument response for {\it Fermi} relevant for the data used in this analysis.  This accounts for the exposure correction and also the finite PSF, for example.  In reality, we expect that the gas-correlated emission is a convolution of the H$_{\rm I}$ map and the cosmic-ray distribution.  While we do not formally account for the latter, the hope is that the spherical-harmonic marginalization can help.  The basic idea is that by fitting the normalization of each harmonic mode of the gas map, the fitting procedure has the flexibility to account for large-scale variations of the gas map (which would arise in actuality from the cosmic-ray diffusion).  In this way, we can reconstruct the cosmic-ray distribution in a data-driven fashion.

We consider adding the \texttt{HI4PI} map to the NPTF analyses in addition to the Model O templates.  We begin by performing purely Poissonian analyses using the standard set of Poissonian templates for Model O (the two Model O templates, the {\it Fermi} bubbles template, the 3FGL PS template, an isotropic template, and the GCE template), but also adding in the H$_{\rm I}$ map with harmonics marginalized up through some $\ell_{\rm max}$.  We perform this analysis in our fiducial ROI ($b > 2^\circ$ and $r < 25^\circ$) with 3FGL sources masked.  We find that the inclusion of the H$_{\rm I}$ map can substantially improve the fit of the model to the data.  For example, the model including the H$_{\rm I}$ map is preferred over that without with a Bayes factor of $2 \ln {\rm BF} \approx 2$, 64, 71, and 81 for $\ell_{\rm max} = 0$, 1, 2, and 3.  The harmonic marginalization makes a significant difference in this case, which is expected since the primary H$_{\rm I}$ map did not have a cosmic-ray spatial morphology already incorporated.  

We then investigate how the inclusion of the H$_{\rm I}$ map affects the results of the NPTF analysis.  We find that including the H$_{\rm I}$ template in the NPTF leads to a slight increase in the evidence for PSs.  For example, including the H$_{\rm I}$ template with harmonic correction up to and including $\ell_{\rm max}$, we find that the evidence in favor of GCE-correlated PSs is approximately $ 2\ln {\rm BF} = $ 22, 21, 19, and 19 for $\ell_{\rm max} = 0$, 1, 2, and 3.  Recall that without the H$_{\rm I}$ template, the Bayes factor in preference for PSs was $ 2\ln {\rm BF} \approx 15$.   Interestingly, this seems to suggest that while the inclusion of the H$_{\rm I}$ template provides a substantially better fit to the data, it has little effect (and, if anything, a positive effect) on the evidence for GCE-correlated PSs.  However, we caution that the H$_{\rm I}$ template and our method of using harmonic marginalization to account for the cosmic-ray morphology is still not leading to fits that describe the {\it Fermi} data at the level of Poisson noise (though it does bring us closer).  We can therefore not rule out the possibility that a better gas-correlated emission map would have a negative effect on the evidence in favor of GCE-correlated PSs.          

\section{Extended Source-count Distribution Results}
\label{app:extended}
\begin{figure*}[t]
\leavevmode
\begin{center}
\includegraphics[width=0.49\textwidth]{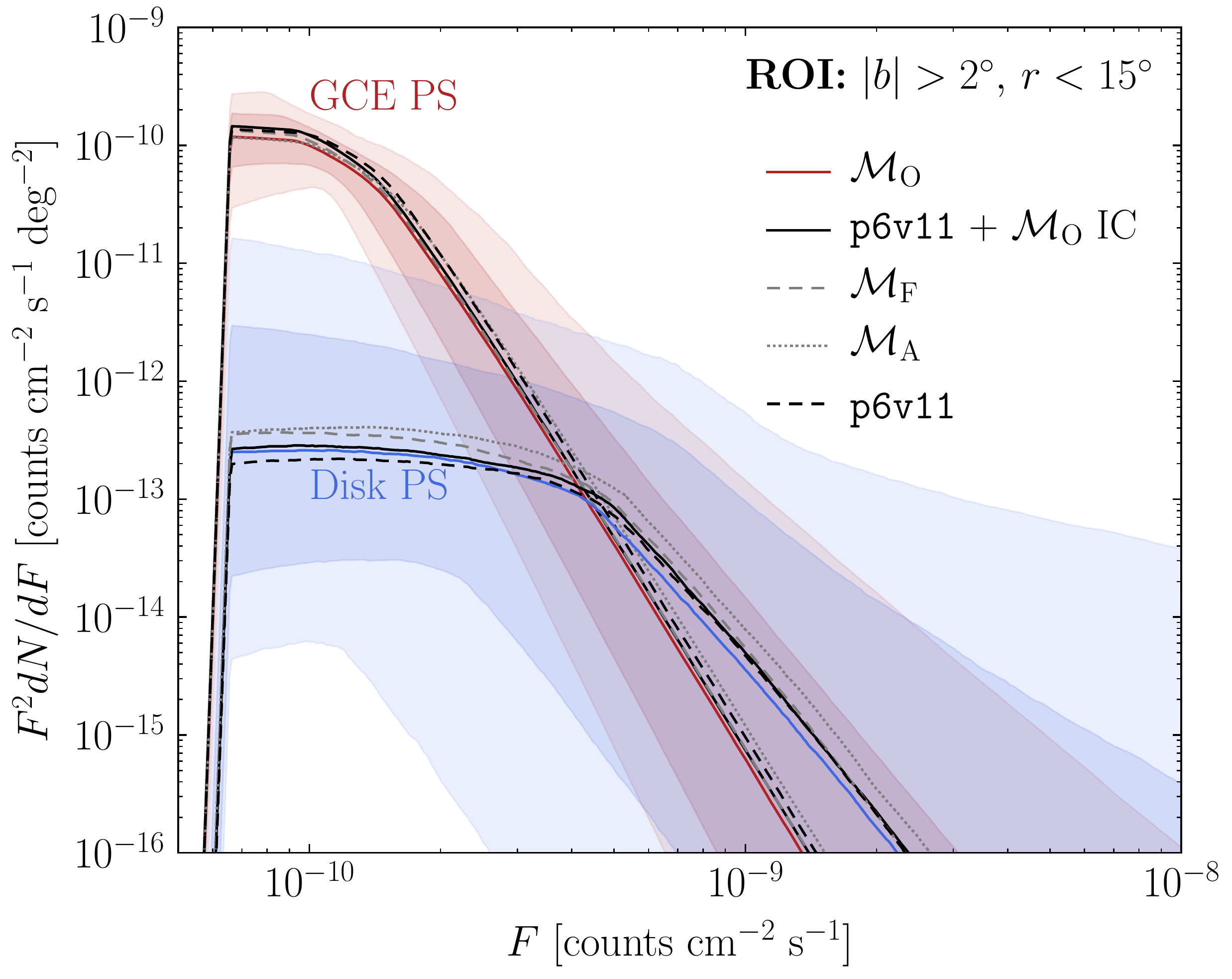} 
\end{center}
\caption{The recovered source-count distributions for the GCE-correlated and disk-corrrelated PSs from the NPTF analyses in the indicated ROI, which is the one that we use to obtain the source-count distribution model parameters for our MC.   We show results for the analyses using different diffuse models, as indicated.  The 68\% containment intervals are from the Model~O analyses.  For the rest of the diffuse models, only the median of the posterior is shown.}
\label{fig:extended}
\end{figure*}

Finally, we provide additional results for the source-count distributions found in the NPTF analyses.  The source-count distributions are summarized in Fig.~\ref{fig:extended} for the ROI $|b| > 2^\circ$ and $r<15^\circ$.  This is the ROI that we use to get the source-count distribution model parameters to generate our MC.

\bibliographystyle{apsrev4-1}
\bibliography{GCE-Harmonics}

\end{document}